\documentclass[12pt,preprint]{aastex}

\usepackage{emulateapj5,apjfonts}
\usepackage{graphicx}
\usepackage{onecolfloat}

\slugcomment{accepted for publication in PASP~~~~~ 
             24/02/2005}

\shorttitle{PMAS I.}
\shortauthors{Roth et al.}

\begin{document}

\twocolumn[    

\title{PMAS, the Potsdam Multi-Aperture Spectrophotometer \\
       I. Design, Manufacture, and Performance}


\author{Martin M. Roth\altaffilmark{1,2}, 
        Andreas Kelz \altaffilmark{1},
        Thomas Fechner,
        Thomas Hahn, \\
        Svend-Marian Bauer,
        Thomas Becker\altaffilmark{1,2}, 
        Petra B\"ohm\altaffilmark{1},
        Lise Christensen\altaffilmark{1},\\
        Frank Dionies,
        Jens Paschke,
        Emil Popow,
        Dieter Wolter}
\affil{Astrophysikalisches Institut Potsdam, An der Sternwarte 16, D-14482 Potsdam, Germany}
\email{mmroth@aip.de}
\author{J\"urgen Schmoll\altaffilmark{1,2}}
\affil{Astronomical Instrument Group, Dept. of Physics, University of Durham, \\
Rochester Buildg., South Road, Durham DH1 3LE, United Kingdom}
\author{Uwe Laux\altaffilmark{3}}
\affil{Landessternwarte Tautenburg, Germany}
\and
\author{Werner Altmann\altaffilmark{3}}
\affil{Konstruktionsb\"uro Altmann, Passau, Germany\\~\\~}

\begin{abstract}
We describe the design, manufacture, commissioning, and performance
of PMAS, the Potsdam Multi-Aperture Spectrophotometer. PMAS is a dedicated
integral field spectrophotometer, optimized to cover the optical
wavelength regime of 0.35--1~$\mu$m. It is based on the lens array -- fiber
bundle principle of operation. The instrument employs an all-refractive
fiber spectrograph, built with CaF$_2$ optics, to provide good
transmission and high image quality over the entire nominal wavelength
range. A set of user-selectable reflective gratings provides low to medium
spectral resolution in first order of approx.\ 1.5, 3.2, and 7~{\AA},
depending on the groove density (1200, 600, 300~gr/mm).
While the standard integral field unit (IFU) uses a 16$\times16$ element 
lens array, which provides seeing-limited sampling in a relatively small 
field-of-view (FOV) in one of three magnifications (8$\times8$, 12$\times12$, or
16$\times16$~arcsec$^2$, respectively), a recently retrofitted bare
fiber bundle IFU (PPak) expands the FOV to a hexagonal
area with a footprint of 65$\times$74~arcsec$^2$. Other special features
include a cryogenic CCD camera for field acquisition and guiding,
a nod-shuffle mode for beam switching and improved sky background subtraction,
and a scanning Fabry-P\'erot etalon in combination with the standard
IFU (PYTHEAS mode). PMAS was initially designed and built as an experimental
traveling instrument with optical interfaces to various telescopes (Calar
Alto 3.5m, ESO-VLT, LBT). It is offered as a common user instrument 
at Calar Alto under contract with MPIA Heidelberg since 2002.
\end{abstract}


]   

\altaffiltext{1}{Visiting Astronomer, German-Spanish Astronomical Centre, Calar Alto, operated by the Max-Planck-Institute for Astronomy, Heidelberg,
jointly with the Spanish National Commission for Astronomy.}
\altaffiltext{2}{Visiting Astronomer, Special Astrophysical Observatory, Selentchuk,
Russia.}
\altaffiltext{3}{Under contract with AIP.}

\section{Introduction}

Unlike optional integral field units which can be deployed in front of a 
conventional slit spectrograph, PMAS, the Potsdam Multi-Aperture
Spectrophotometer, was designed as a dedicated integral field  spectrophotometer. 
It was built entirely at the Astrophysical Institute Potsdam (AIP), Germany. 
Initially designed as a travelling instrument, PMAS was commissioned and 
subsequently installed as a common user instrument at the Cassegrain focus 
of the Calar Alto  Observatory 3.5m Telescope in southern Spain. 

The instrument was specifically designed to address the science case of
3D spectrophotometry of spatially resolved, individual objects, with an 
emphasis on broad wavelength coverage in the optical wavelength regime. 
The spectrograph camera was designed to accommodate a single 2K$\times$4K /
15$\mu$m pixel CCD, or a 2$\times$2K$\times$4K mosaic for the full
field, providing 2048 or 4096 pixels in the spectral direction,
and 4096 pixels in the spatial direction, respectively. 
The standard  lens array IFU has a relatively small 
field-of-view (FOV) of 16$\times$16 square spatial elements (``spaxels'') on the sky, 
corresponding to 8$\times$8~arcsec$^2$ with seeing-limited sampling of 0.5~arcsec/spaxel,
or 16$\times$16~arcsec$^2$ in the 2$\times$ magnification mode with
a sampling of 1.0~arcsec/spaxel. The instrument is based on a modular design,
including a telescope module with fore optics and the standard IFU,
a fiber module to couple the light from the IFU to the spectrograph, 
and the fiber spectrograph with CCD camera. The fore optics, standard IFU, 
and spectrograph optics are all built from fused silica, CaF$_2$, or other 
UV-transparent media such that PMAS is optically corrected and transparent 
from 0.35---1~$\mu$m. The instrument also includes a direct imaging cryogenic 
CCD camera for field acquisition and guiding.

\begin{table*}
 \begin{center}
 \caption{Main Instrument Parameters\label{TBL-1}}
 \begin{tabular}{ll}
 \tableline\tableline
SPECTROGRAPH:            &  \\
\tableline
focal station:           & cassegrain  \\
principle of operation:  & IFU + fiber-coupled spectrograph  \\
spectrograph type:       & fully refractive f/3 collimator and f/1.5 camera  \\
wavelength range:        & 0.35-1.0~$\mu$m (CaF$_2$ optics)  \\
gratings:                & 1200, 600 and 300 gr/mm reflective gratings  \\
linear dispersion:       & approx. 0.35, 0.8, and 1.7 {\AA}/pixel, respectively  \\
spectrograph FOV:        & 60 $\times$ 60 mm$^2$  \\
spectrograph detector    & single 2K$\times$4K, or 2$\times$~2K$\times$4K
                           mosaic CCD camera, 15$\mu$m pixels \\ 
   &   \\
LENS ARRAY IFU (LARR) &  \\
 \tableline
principle of operation:  & square lens array with magnifying fore optics  \\
lens array:              & 16 $\times$ 16 square elements, 1mm pitch \\ 
fore optics magnifications: & --- 0.5 arcsec sampling, 8 $\times$ 8 arcsec$^2$ FOV \\
                         & --- 0.75 arcsec sampling, 12 $\times$ 12 arcsec$^2$ FOV \\
                         & --- 1.0 arcsec sampling, 16 $\times$ 16 arcsec$^2$ FOV \\
fiber configuration:     & 256 OH-doped fibers, 100~$\mu$m core diameter \\
filters:                 & filter slider with five 1-inch round filter positions \\ 
  &  \\
FIBER BUNDLE IFU (PPAK) &  \\
 \tableline
principle of operation:  & focal reducer + hexagonal packed fiber-bundle \\
fore optics:              & focal reducer lens F10 to F/3.3, plate scale: 17.7~arcsec/mm \\ 
fiber configuration:     & 331 object + 36 sky + 15 calibration fibers, 150~$\mu$m core diameter \\
field-of-view:           & 74 $\times$ 65 arcsec$^2$, hexagonal \\
spatial sampling:        & 2.7 arcsec per fiber diameter, 3.45~arcsec pitch
                           (nearest neighbour) \\ 
filters:                 & user-selectable 35$\times$140~mm$^2$ order separating filter in \\
                         & spectrograph collimator mount;   \\
                         & alternatively: 50~mm round filter in front of fiber bundle\\  
     &   \\
ACQUISITION AND GUIDING CAMERA   &  \\
 \tableline
detector:                & 1K$\times$1K CCD, thinned backside illuminated, AR coated \\
field-of-view:           & 205 $\times$ 205 arcsec$^2$ \\
pixelsize, scale:        & 24$\mu$m, 0.2~arcsec/pixel \\
filters:                 & filter slider with four 2-inch round filter positions \\ 
 \tableline \tableline
 \end{tabular}
 \end{center}
 \end{table*}

\clearpage

\begin{figure*}[htb]
\centering\includegraphics[width=0.8\linewidth,clip]{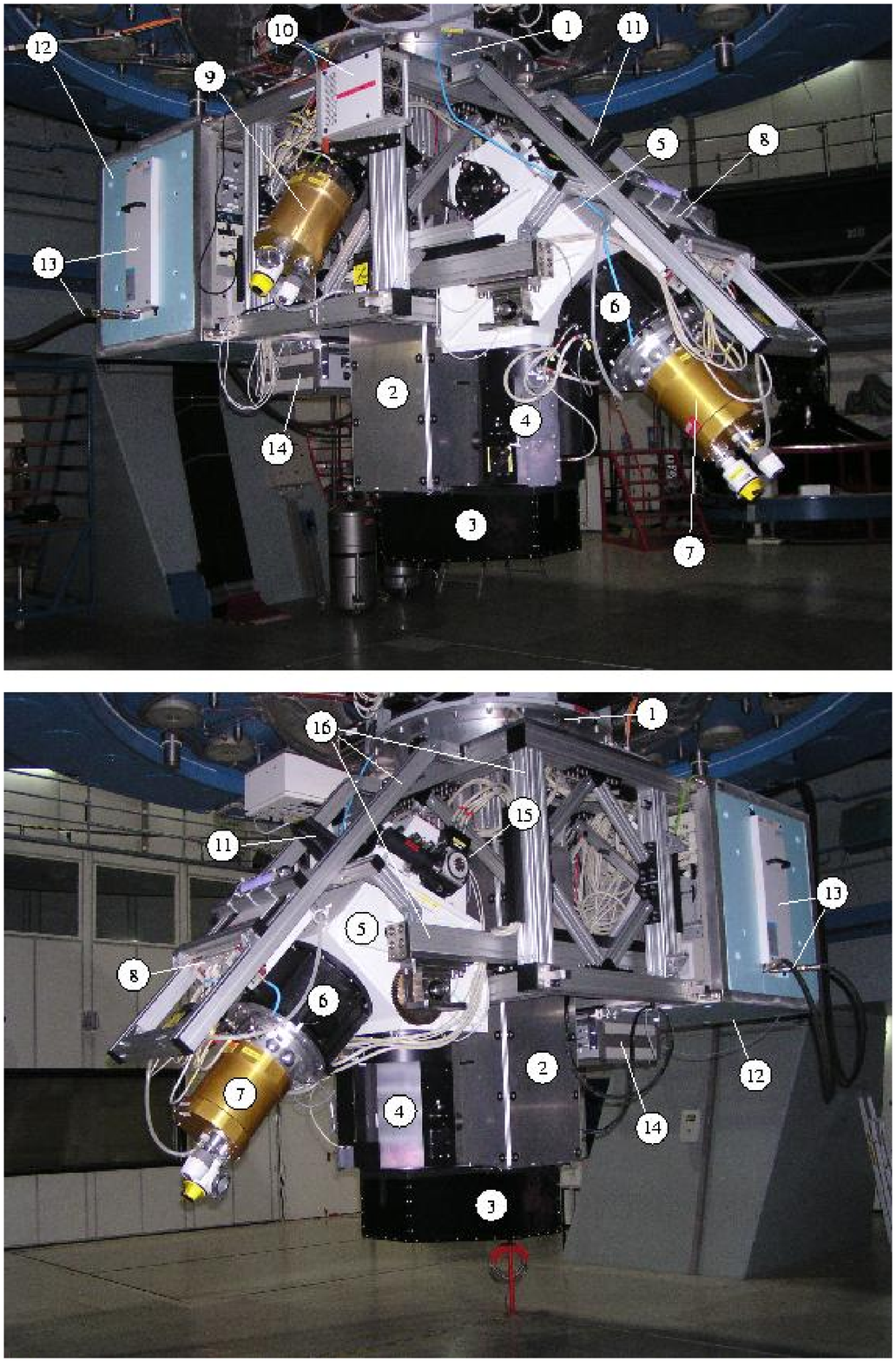}
\figcaption{PMAS at the Calar Alto 3.5m Telescope Cassegrain Focus :
(1) Instrument Flange, (2) Telescope Module, (3) Fiber Module,
(4) Spectrograph Collimator, (5) Spectrograph Center Case, 
(6) Spectrograph Camera, (7) Spectrograph CCD, (8) Spectrograph CCD Controller, 
(9) Acquisition and Guiding (A\&G) CCD, (10) A\&G CCD Controller,
(11) Grating Platform, (12) Electronics, (13) Cooling Subsystem,
(14) Fabry-P\'erot Controller, (15) Grating Rotator,
(16) Main Frame. The instrument measures approximately 2m from the
     bottom to the telescope flange.
\label{OVERVIEWPIC}}
\end{figure*}

\clearpage

Since commissioning at Calar Alto in 2001, and facilitated by the open and modular
structure of PMAS, several new features have been added to the initial 
configuration of the instrument:  an off-axis fiber bundle
(``{\em PPak}'')  for an enlarged FOV$>$1~arcmin (Verheijen et al.\ 2004),
a {\em nod-shuffle} mode of operation for faint object integral field 
spectroscopy (Roth et al.\ 2002d),
and the {\em PYTHEAS} mode with a scanning Fabry-P\'erot etalon for an increased
spectral resolution over a large free spectral range (Le Coarer 1995).
We shall give detailed reports on these special modes in future papers II., III.,
and IV., respectively.

The initial conceptual design, progress reports on  assembly, integration,
and test, and first results from commissioning are given in Roth et al.~1997, 
Roth et al.~1998, Roth et al.~2000a,b, Roth et al.~2003, Kelz et
al.~2003a, and Kelz et al.~2003b. First results from a science verification 
run were reported in Roth et al.~2004a. 

In this paper, we briefly outline the science drivers for the instrument 
(\S~\ref{SCIENCECASE}), describe the final design, manufacture, and commissioning
(\S~\ref{INSTRUMENT}), and provide information on the performance at the 
telescope (\S~\ref{sec:PERFORM}).

\section{Science Case and Requirements}
\label{SCIENCECASE}
Integral field spectroscopy\footnote{also referred to as 
{\em 3D spectroscopy}} is an emerging observing technique on the
verge of becoming a standard method. After a period of experimental
developments since the late 1980's, several facility instruments in the
optical and NIR were planned and subsequently built for some 4-8m class telescopes,
e.g.\ INTEGRAL/WHT (Arribas et al.\ 1998), GMOS-IFU/Gemini (Allington-Smith et
al.\ 2002), GNIRS-IFU/Gemini (Elias et al.\ 1998), FLAMES-IFU/VLT (Pasquini et al.\ 2000), 
VIMOS-IFU/VLT (Le Fevre et al.\ 2003), SINFONI/VLT (Eisenhauer et al.\ 2003).
In this ground-breaking era, PMAS was intended to make a contribution to the
development of {\em crowded field 3D spectroscopy}, analogous to
the successful introduction of {\em crowded field photometry}
with the advent of CCD detectors in Astronomy. 

Unlike most of the 
more conventional applications of 3D spectroscopy, where the 
simultaneous coverage of a -- preferably large -- {\em area} for 
the  purpose of creating 2-dimensional maps of emission line intensities, 
velocity fields,  velocity dispersions, absorption line indices, etc.\ 
have determined the technical requirements (e.g.\ de Zeeuw et al.\ 2002), 
it was the primary goal for PMAS to provide an optimal 
{\em sampling of the point-spread-function} (PSF) over the {\em entire
optical wavelength region}, from the atmospheric cutoff in the UV
to the NIR. Such properties would make an integral field spectrograph
an ideal tool for the spectrophotometry of faint point sources, with
superior properties over slit spectrographs in terms of slit losses,
of sensitivity to differential atmospheric refraction and pointing errors,
and of accurate background subtraction in fields where source confusion
is an issue (Roth et al.\ 1997, 1998, 2000c). 

Among the major science cases to demand such properties there were
considered: spectrophotometry of resolved stellar populations in nearby 
galaxies (including extragalactic planetary nebulae, H$\;$II regions, 
optical counterparts of  ultraluminous X-ray sources), supernovae, 
gravitationally lensed QSOs, QSO host galaxies, high redshift galaxies. 
The recent discussion of the science case for extremely large telescopes 
(ELTs) has shown that this motivation is, in fact, timely and relevant for 
future developments of ground-based optical Astronomy (e.g.\ 
Najita \& Strom 2002, Hawarden et al. 2003). We shall summarize some first
PMAS results on these topics in \S~\ref{sec:CAHA}.

The PMAS requirements derived from these science cases include wide
wavelength coverage of 0.35--1~$\mu$m, high efficiency, low to medium spectral 
resolution, seeing-limited spatial sampling at a modest FOV, high stability,
a supporting direct imager for field acquisition, guiding, and differential
spectrophotometry, and a self-contained design (traveling instrument). 

The science case for the bare fiber bundle IFU (PPak) is identical to the one of SparsePak 
(Bershady et al.\ 2004), see Verheijen et al.\ 2004. It is complementary
to the standard IFU in the sense that it adresses the problem of low surface
brightness 3D spectroscopy using large spaxels over a large FOV at the expense
of spatial resolution. 

\section{Instrument Description}
\label{INSTRUMENT}

\subsection{Overview}

PMAS is based on the principle of a fiber-coupled lens array IFU, 
employing a dedicated fiber spectrograph. Initially,
PMAS was required to operate as a traveling instrument. Therefore,
the instrument was designed for operation at the cassegrain 
focus station of a 4-8m class telescope with a self-contained structure,
having a minimal number of interfaces to its environment. It has a modular 
layout with the following major components (Fig.~\ref{OVERVIEWPIC}): 
telescope module with flange to the telescope, the main frame as a support 
structure for the remaining modules, the fiber spectrograph, the fiber module, 
connecting the telescope module and the fiber spectrograph, and the electronics 
rack for the instrument control subsystem. Except for the telescope flange,
the 230VAC power cable, a connection to the LAN, and two pairs of cooling hoses 
for a closed cycle heat exchanger, PMAS has no connections to the outside
world.

\subsection{The Spectrograph Module}

In terms of importance and costs, the fiber spectrograph is the dominant 
component of the instrument and determines to a large extent its overall
properties. It employs an all-refractive optical
design with an f/3 collimator, a beam size of 150mm in diameter, 
interchangeable reflective gratings, and an f/1.5 camera with a
corrected focal plane of 60$\times$60~mm$^2$. The nominal wavelength
range is 0.35--1~$\mu$m.

\begin{figure}
\plotone{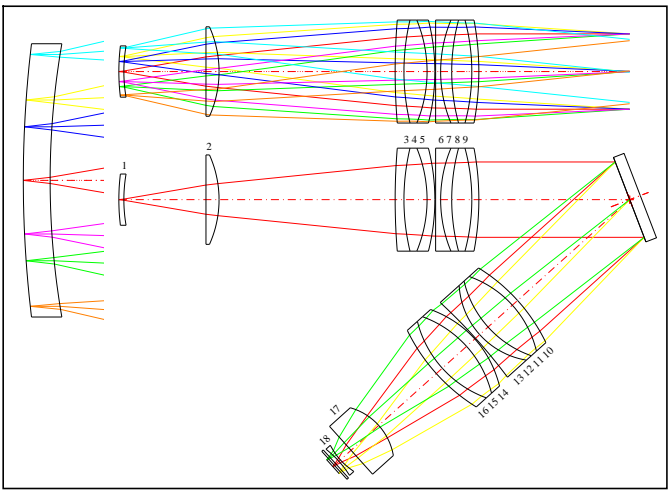}
\caption{Bottom: Cross-section of the fiber spectrograph optical system in
         the plane of dispersion. Top: collimator cross-section,
         perpendicular to the dispersion. 
         Insert to the left: magnified detail of
         the fiber coupling lens (from the collimator graph at the top).
         The lens materials are as follows: quartz: lens 1, 2, 17; CaF$_2$:
         lens 4, 8, 11, 15; BaK4: lens 3, 5, 7, 9, 10, 12, 14, 16, 18; 
         LF5: lens 6, 13. The last element in the optical train is the
         plane-parallel quartz window of the CCD cryostat.
         \label{FPSPEC_TOTAL}}
\end{figure}

The optical design principle and optimization was performed by
UL with proprietary software (Laux 1999). The result was double-checked 
using the commercial ZEMAX and Code-V programmes.
The collimator and camera systems are characterized by a similar 
layout (see Fig.~\ref{FPSPEC_TOTAL}), each consisting of the combination of two
lens groups (triplet, quadruplet), combined with two singlet
field lenses near the fiber slit (collimator) and the detector focal plane 
(camera), respectively. The camera has two aspherical surfaces, viz.\ the 
first surface of the triplet (lens 14), and the first surface of the field 
lens (lens 18), as seen from the grating. 
The camera field lens is followed by a plane-parallel quartz
plate, acting as the cryostat window of the CCD detector subsystem.
The external focus of the camera has the significant advantage of allowing
for an easy interchange of detectors.

As a result of the compact configuration of the two front lens groups,
the optical system presents comparatively few (18) glass-air interfaces
which are all covered with high quality broadband AR coatings, yielding an
average residual reflectance of 0.8\% per surface in the wavelength
interval 0.35--1~$\mu$m. Due to the excellent transmission properties 
of the glasses, the system has a high throughput down to 350nm. 
The image quality is well-matched to the nominal pixel 
size of 15~$\mu$m (Tab.~4). There is a modest amount of vignetting 
($~\approx25\%$) near the edge of the spectrograph FOV. The masses of the
complete collimator and camera subsystems, including the massive 
lens mounts, are 60~kg each, respectively. The optical systems
were manufactured and assembled by Carl Zeiss Jena, Germany. The
results of the acceptance tests were reported by Roth et al.~2000b.

The spectrograph mechanical design is characterized by a solid aluminum cast
housing, which keeps the heavy collimator and camera optics in place.
Both lens barrels are protected by stiff aluminum tubes which are
mounted to the center case, but decoupled from the lenses. The
camera tube carries the spectrograph CCD camera, which is also mechanically
decoupled from the optics. The design was optimized for high 
mechanical stability using the ANSYS finite element code (Dionies 1998).
The whole spectrograph subsystem is suspended at two points near the
center of gravity and mounted to the main frame such as to avoid the
variation of torque over different pointings of the telescope.

The exchangeable reflective grating is mounted on a 360$^\circ$
rotator which accommodates any required grating orientation, depending
on the choice of ruling and blaze. The gratings are fixed in a 6-point
kinematic mount, the blazed front surface coinciding with the rotation
axis of the device. The grating rotator is balanced by counter-weights, 
again to avoid any torque effects under varying orientations.

PMAS is equipped with a set of interchangeable reflective gratings (Tab.~2),
each of which is mounted in an individual frame and stored in a protective
cartridge when not installed in the instrument. The gratings were
manufactured by Richardson Grating Laboratory, Rochester, U.S.A., as
replicas from master gratings for astronomical use with a ruled area
of 154$\times$206~mm$^2$ and a blank size of 165$\times$220$\times$35~mm$^3$.
Each grating frame carries a 4-bit code made out of small magnets (grating ID),
which is read upon loading a grating from its cartridge into the
grating rotator and is used to generate a corresponding FITS header
entry ~{\tt GRAT\_ID}~ for each spectro\-graph data file. 

The spectral resolution depends on the choice of grating, the 
zero-th order projected fiber size on the grating (4 pixels or
6 pixels diameter for the lens array and PPak IFUs, respectively),
anamorphic (de-)magnification (Schweizer 1979), and tilt and orientation 
of the grating. Depending on the orientation, i.e.\ the normal of the
blaze surface facing either the camera (``{\em forward}''), or the collimator
(``{\em backward}''), the anamorphism is effected in opposite ways, and
the tilt must be adjusted accordingly. The grating cartridges are built
such as to allow operation both in forward or backward blaze 
with just a minor mechanical modification at the frame. The default
orientation of PMAS gratings for spectrophotometry is {\em forward}. 
However, a significant increase in spectral resolution is obtained 
for the J1200 grating in 2nd order backward blaze, as discussed in paper~II.
A comprehensive listing of grating parameters as a function of grating
angle and orientation are given in the PMAS online 
manual\footnote{\url{http://www.aip.de/groups/opti/pmas/OptI\_pmas.html}}.

\vskip 5mm
\centerline{TABLE 2}
\vskip 1mm
\begin{tabular}{lrrcrrr}
Grating &  ID &  lpmm  &   D &  $\alpha_{blaze}$ & $\lambda_{blaze}$ &   $\Delta\lambda$  \\  
\hline\hline
U1200  & 1   & 1200   & 0.39   & 10.4  &  300    &  794 \\
V1200  & 2   & 1200   & 0.35   & 17.5  &  500    &  725 \\
R1200  & 3   & 1200   & 0.30   & 26.7  &  750    &  609 \\
I1200  & 4   & 1200   & 0.22   & 36.8  & 1000    &  460 \\
J1200  & 5   & 1200   & 0.22   & 46.0  & 1200    &  450 \\
J1200$^{(*)}$  & 5   & 1200   & 0.17   & 46.0  &  600    &  341 \\
~U600   & 6   &  600   & 0.81   &  5.2  &  300    & 1656 \\
~V600   & 7   &  600   & 0.80   &  8.6  &  500    & 1630 \\
~R600   & 8   &  600   & 0.75   & 13.9  &  800    & 1533 \\
~U300   & 9   &  300   & 1.67   &  2.5  &  300    & 3404 \\
~V300   & 10  &  300   & 1.67   &  4.3  &  500    & 3404 \\
\hline
\end{tabular}
\vskip 2mm
\noindent
{\footnotesize {\em PMAS grating parameters.
Column~1: name, Col.2: identifier, Col.3: groove density [gr/mm],
Col.4: reciprocal dispersion [{\AA}/pixel], 
Col.5: blaze angle, Col.6: blaze wavelength [nm],
Col.7: wavelength coverage for 2K$\times$4K CCD [{\AA}]. 
Note $^{(*)}$: 2nd order backward.}}
\vskip 7mm

\subsection{Telescope Module}
\label{sec:TELMOD}

The telescope module is designed for three major purposes: to
re-image the telescope focal plane onto the lens array (fore optics), 
to illuminate the lens array from an internal calibration light source
(calibration unit), and to project an area around the IFU onto a camera for 
target acquisition and guiding (A\&G subsystem). For the latter, either an
external slit-viewing TV guider camera can be used, or, alternatively, 
a cryogenic CCD camera which is internal to the PMAS instrument.

\begin{figure}[t!]
\vskip 5mm
\centering\includegraphics[width=1.0\linewidth,clip]{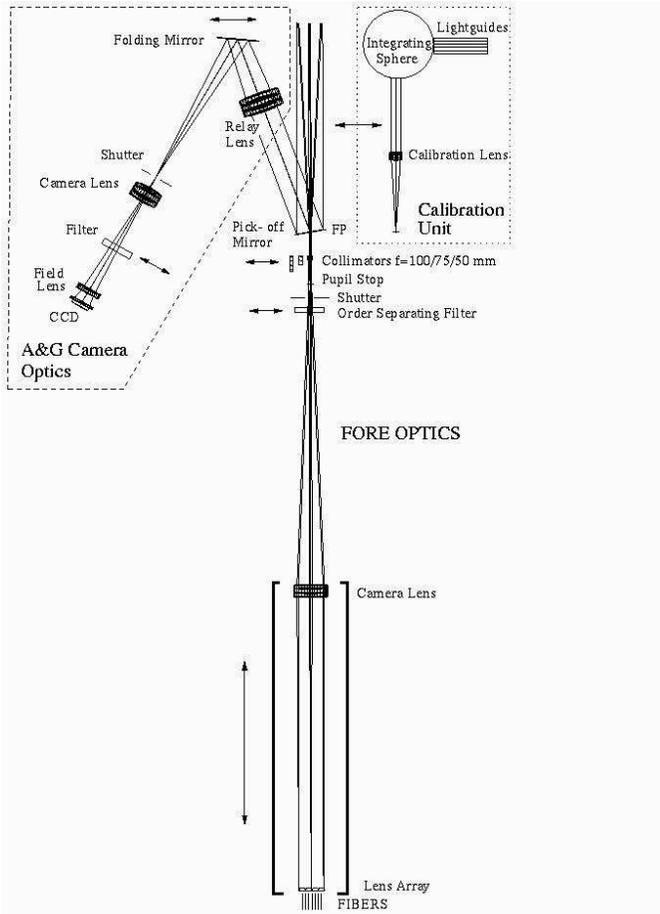}
\vskip 2mm
\figcaption{Cross-section of the three optical systems of the PMAS 
   telescope module: fore optics, calibration unit, and
   A\&G camera system.
    \label{TMOPTICS}}
\end{figure}

The three optical systems are shown schematically in Fig.~\ref{TMOPTICS}.
Light entering from the telescope is combined in the focal plane FP (intersection
of the tilted A\&G pick-off mirror and optical axis, upper part of the drawing). 

The optical train of the fore optics begins at a central hole in the pick-off mirror
which gives way for the field-of-view of the lens array, 
onto which the telescope focal plane is imaged with a magnification of 1:11.8 
(for the standard sampling of 0.5 arcsec per lens). 
Re-imaging is accomplished through the combination of a 
collimating lens, and a camera objective.
Owing to a double 2$\times$f arrangement of these lenses, an intermediate
pupil is formed at the conjugate of the telescope focal plane. For unbaffled
telescopes this is a convenient location for a Lyot stop. Near
the intermediate pupil there is also a shutter, and a linear
stage, accommodating up to 5 order separation filters.
The double 2$\times$f scheme has the disadvantage of a relatively
long optical train ($\approx$1m), but the significant advantage
of allowing for future experimental upgrades, e.g.\ Fabry-P\'erot etalon 
(PHYTEAS mode, Le Coarer 1995, see paper IV), or polarimetry. Also,
telecentricity for optimal lens array--fiber coupling is ensured without the need 
for an additional field lens in front of the IFU.
Three different magnifications are chosen by interchangeable collimator lenses
which are mounted on a linear stage. Tab.~3 lists the parameters for lenses
with focal length 50mm, 75mm, and 100mm, respectively, which were computed for three 
telescopes (Calar Alto 3.5m, VLT, LBT f/15 Gregorian focus). 
Switching to another scale requires also to adjust the position of
the following camera lens and the lens array along the optical axis of the
fore optics.
They are mounted at the ends of a stable tube, thus forming a single unit 
whose position along the optical axis can be adjusted by translation 
on a precision linear stage.

The calibration unit consists of an 
integrating sphere, whose exit port is imaged onto the intermediate pupil 
through the calibration lens and the fore optics collimator,
imitating as closely as possible the telescope beam. The correct f-ratio 
is defined by the calibration lens, while the telescope pupil is mimicked with a stop 
at the output of the integrating sphere, including a central obscuration 
to simulate the M2 shadow.
The calibration unit is mounted on a linear stage and is moved into the beam for
calibration exposures.

\begin{figure}[t!]
\vskip 3mm
\centering\includegraphics[width=1.0\linewidth,clip]{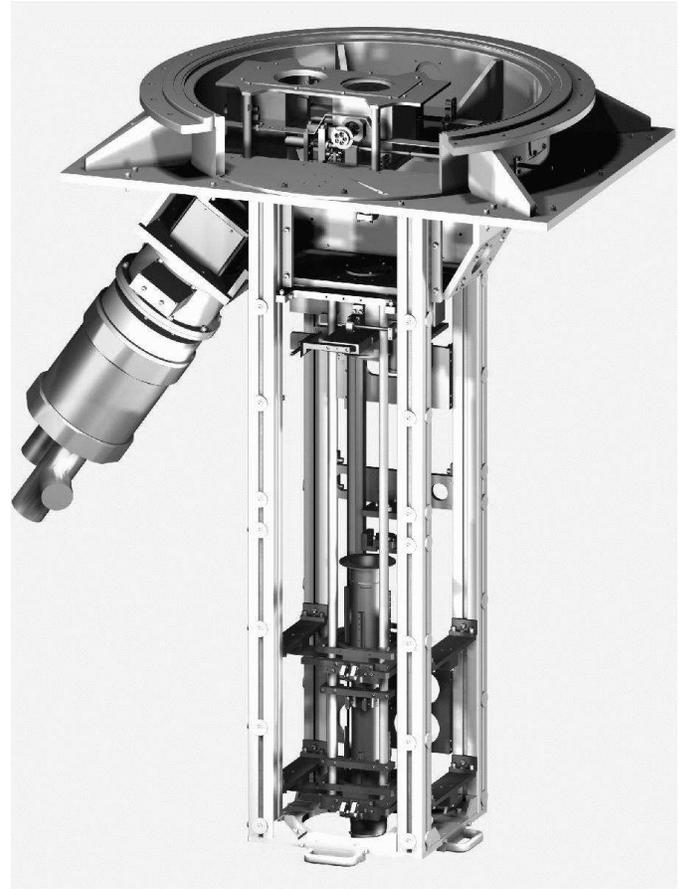}
\vskip 5mm
\caption{Mechanical layout of the Telescope Module, shown without
         cover plates (CAD drawing). The break-out in the telescope
         flange provides insight into the linear stage with the
         calibration unit and A\&G camera subsystem 
         devices. \label{TELMOD_CAD}}
\end{figure}

The A\&G camera subsystem 
is similar to a conventional slit-viewing system where the
reflective frontside of a spectrograph slit is observed with a TV guiding
camera. In the case of PMAS, the slit is replaced by the tilted pick-off
mirror. The combination of a relay lens, a folding mirror, a camera lens, and a 
field lens projects the plane of the pick-off mirror onto a CCD, 
which is used for target acquisition and offset guiding. A filter stage mechanism 
accommodates up to four 50mm filters at a location roughly half-way between 
the camera and the field lens. 

Mechanically, the telescope module consists of a rigid aluminum-cast housing with
a flange to the telescope, forming also the base for the support frame, 
and a tower-like structure extending from the flange down to the fiber
module (Fig.~\ref{TELMOD_CAD}). 
The tower acts as a stable mechanical support for an
optical bench which is carrying the entire fore optics system. It is essentially
self-contained and can be dismounted for shipping as a whole subsystem.
The optical bench consists out of LINOS Macro Bench elements and custom
made components for the movable parts, namely the linear stages for the
collimator lenses, order separating filters, and the camera--lens array tube. 

The calibration unit on its linear stage is visible in Fig.~\ref{TELMOD_CAD}
through the break-out of the flange. An input piece provides a port for
six waveguides simultaneously to feed light from remote light sources
(continuum and spectral line lamps) into the integrating sphere.

The A\&G camera is seen on the left-hand side of Fig.~\ref{TELMOD_CAD}
as the prominent LN$_2$ dewar, attached to the telescope module housing at an angle.
Its flange, together with another Macro Bench assembly, forms a modular
subsystem which carries all opto-mechanical elements, and
which can be easily removed from the housing for inspection and maintenance.
The cap of the filter stage, which itself is yet another modular subsystem, is
visible as a rectangular structure at the front part of the detector flange.

\vskip 5mm
\centerline{TABLE 3}
\vskip 1mm
\centerline{\begin{tabular}{ccccc}
f $'_{Coll}$ [mm] & $\Gamma$ & IP [mm] & $\mu$P [$\mu$m] & pitch [$''$]\\ 
\hline\hline
                 &                 &         &        &     \\
\multicolumn{5}{c}{Calar Alto RCC 3500/35000, ~~~f/10}\\
\hline
    50.0~~   &  ~700.0~ &  5.00 & 42.6 & 0.50\\
    75.0~~   &  ~466.7~ &  7.50 & 63.8 & 0.75\\
    100.0~~  &  ~350.0~ & 10.00 & 85.1 & 1.00\\
             &                 &         &        &     \\
\multicolumn{5}{c}{VLT 8115/108825,  ~~~f/13.4}\\
\hline
    50.0~~   & ~2176.5~ &  3.73 & 31.7 & 0.16\\
    75.0~~   & ~1451.0~ &  5.59 & 47.6 & 0.24\\
   100.0~~   & ~1088.3~ &  7.46 & 63.5 & 0.32\\
                 &                 &         &        &     \\
\multicolumn{5}{c}{LBT 8408/123765,  ~~~f/14.7}\\
\hline   
   50.0~~    & ~2475.3~ &  3.40 & 28.9 & 0.14\\
   75.0~~    & ~1650.2~ &  5.10 & 43.4 & 0.21\\
   100.0~~   & ~1237.6~ &  6.79 & 57.8 & 0.28\\
\end{tabular}}
\noindent
\vskip 2mm
\noindent
{\footnotesize {\em Fore optics parameters. Columns 1 through 5:
collimator focal length, intermediate pupil demagnification factor
$\Gamma$, diameter of intermediate pupil, nominal diameter of micropupil (fiber
input), projected lens array pitch on the sky.}}
\vskip 5mm

\subsection{Integral Field Units}

\subsubsection{Lens Array IFU (LARR)}
\label{sec:LARRIFU}

The main IFU of PMAS for the purpose of performing spatially resolved 
spectrophotometry is the lens array, mounted at the end of the fore optics 
optical train. The open modular design allows
for the exchange with another device without necessarily having to
modify other major subsystems. The current lens array is a monolithic
16$\times$16 elements, 1mm pitch square array with square lenslets,
made from fused silica. It was manufactured by Advanced Microoptics
Systems GmbH, Saarbr\"ucken, Germany, as a custom-design element with 
aspherical lenslet front surfaces, and a common plane backside where the fiber
bundle is attached (Roth et al.~2000a). The optimal surface quality derived
from an interferometric acceptance test was measured to 50~nm r.m.s.,
excluding a strip of approx.~10$\mu$m width at the four intersections of a
surface with its four nearest neighbors. Owing to the aspherical surface,
the lens is, in principle, free of spherical aberration. However, due
to random surface defects and an edge effect at the borders of a lenslet,
which cannot be made arbitrarily sharp, the overall real image quality
is less than ideal (see  \S~\ref{sec:LARRTEST}). The output F-number
is 4.5. Details of the coupling of the lens array to the fiber bundle are 
described in \S~\ref{sec:FIBERARRAYS}. 
\vskip 5mm

\begin{figure}[t!]
\plotone{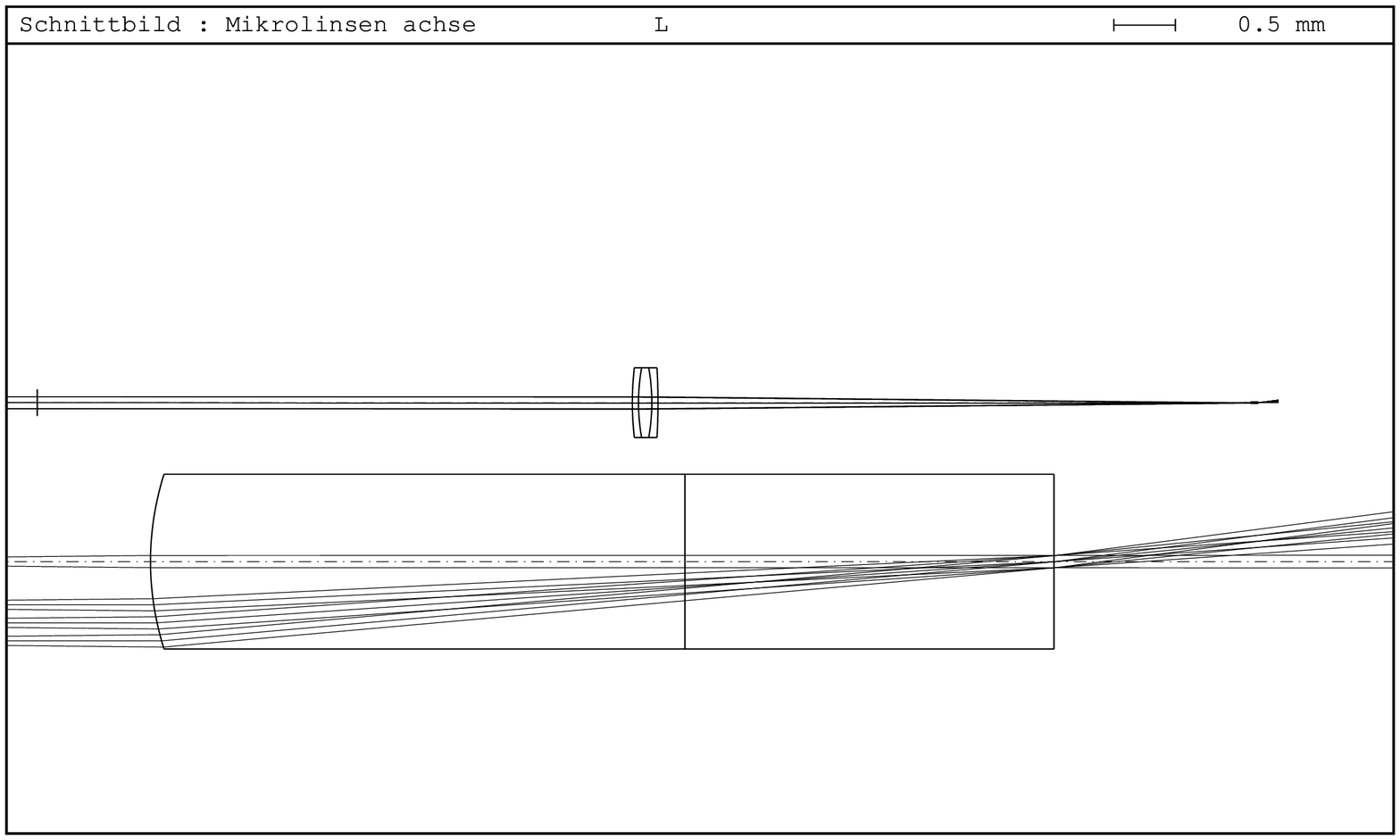}
\vskip 1mm
\plotone{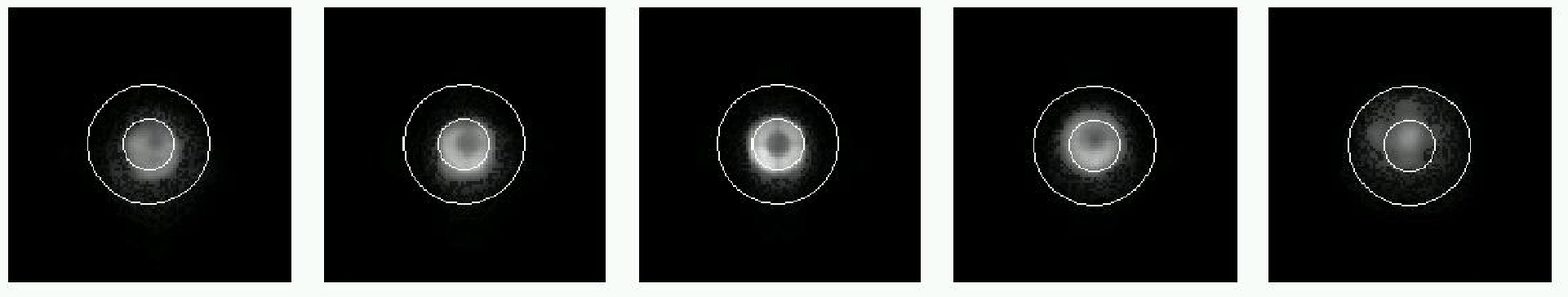}
\caption{Lens array pupil imagery. Upper panel: cross section
of lens array with fore optics (top: pupil -- camera lens -- lens array,
bottom: magnified view of a single lenslet). Lower panel:
through-focus series of micropupil image at the backside of the
lens array,
observed with microscope + CCD imager in the lab. The in-focus,
sharp frame (middle) shows the telescope pupil with M2 shadow 
and spider, created with telescope simulator optics; inner circle:
nominal micropupil diameter (43$\mu$m), outer circle: fiber core
diameter (100$\mu$m).
         \label{PUPILIMAGE}}
\end{figure}

\subsubsection{Bare Fiber Bundle IFU (PPak)}
\label{sec:PPAK}

In addition to the standard IFU using the on-axis lens array, PMAS was equipped 
with a second IFU in 2004. This IFU consists of a bare fiber bundle (called PPak), 
which is placed approximately 6 arcminutes off-axis, as not to obstruct the field 
for the lens array
nor the A\&G camera.  The main purpose of PPak is (1) to provide a 
wide field-of-view and (2) high light collecting power per spectra, rather than 
contiguous spatial sampling. A focal reducer lens is placed immediately in 
front of the fiber bundle, to reduce the telescope beam from F/10 to F/3 
and to change the telescope plate scale from 5.9$''/$mm to 17.8$''/$mm. 
Altogether the PPak-bundle features 382 fibers, of which 331 are placed in a 
densest-packed hexagonal array that projects to $74'' \times 65''$ on the sky. 
36 Sky fibers are arranged within six surrounding mini-IFUs, located $72''$ 
away from the central object fiber. 
Finally, 15 fibers are diverted from the IFU and can be illuminated by calibration 
lamps, to provide simultaneous calibration spectra within the science exposures.      

The combination of individual fiber sizes of 2$''$.7 with the high efficiency 
and wavelength coverage of the PMAS spectrograph, makes PPak a unique 
tool to study extended low-surface brightness objects,  which require high
light collecting power and a large FOV. A more detailed description is given
paper~II and Kelz et al.\ 2004.

\subsection{Fiber Module}

The purpose of the fiber module is to rearrange the
2-dimens\-ional information, sampled by the IFU, onto a linear pseudo-slit.
Additionally, the fibers decouple the IFU and fore optics from the spectrograph,
with the benefits that any mechanical flexure within the telescope module is not
transferred to the spectrograph, and seeing effects do not change the
spectral re\-so\-lution (i.e. the slit width).

PMAS is used for the spectroscopy of crowded fields and of faint,
background-limited objects. For these targets poor flatfield calibration
is a major limitation for the extraction of reliable spectrophotometry.
PMAS features a novel design of exchangeable fiber arrays. Instead of a
permanent bound, both ends of the PMAS fibers are mechanically aligned
and optically matched to lenses in such a way, that fibers can be replaced.
Firstly, this allows the exchange of any fibers that perform below average,
until the overall efficiency is maximized and discrepancies between
individual spectra are minimized. Secondly, it makes it possible to
maintain the performance of the fiber unit during its lifetime, as any
damaged or broken fibers can be replaced without the need of building
a completely new IFU.
While this adds certain complications to the manufacturing process,
(small mechanical devices, potential damage of the fiber end-faces
during assembly, imperfections of the index-matching, alignment)
these can be addressed by a strict quality control.
The aim of the PMAS fiber module is to achieve high efficiency while ensuring a
photometric stable flatfield characteristic over time.

\label{sect:FIBMOD-design}
\begin{figure}[t!]
\vskip 3mm
\hspace{-15mm}
\includegraphics[scale=0.38,angle=-90]{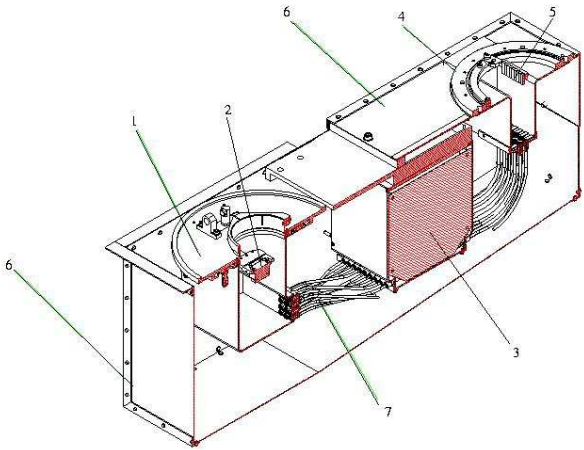}
\vskip 3mm
\caption{Cut view of the PMAS fiber module with components:
        1=IFU holder, 2=lens array, 3=fiber loop box, 
        4=fiber slit, 5=slitlets, 6=fiber module cover,
        7=fiber tubing, (not shown are the 256 individual fibers)}
\label{fig:fibmod_cut}
\end{figure}

As PMAS was specified to be a medium-resolution spectrophotometer,
it was decided to avoid a bench-mounted remote spectrograph with long 
fiber lengths, which would be subject to movement and create modal 
noise (Schmoll et al.\ 2003).
Instead, PMAS features fiber lengths of approximately 2 meters.
This ensures that the internal attenuation of the fibers is low ($\sim 98\%$),
while the input modes are still sufficiently scrambled.

The fiber module consists of three sub-units:
the integral field section, the fiber loop box, and the fiber slit.
There is no stiff mechanical connection between them,
therefore no flexure can be transmitted from one end to the other.
Both the individual units, and the overall assembly are fully enclosed by
a cover for the purpose of mechanical protection and light-tightness
(see Fig.~ \ref{fig:fibmod_cut}).

\label{sec:FIBERARRAYS}
The fibers chosen are FVP100120140 from Polymicro Inc., whose core size, FRD 
performance and spectral attenuation were found to best fit the PMAS requirements.
These step-index silica fibers are OH enhanced,
have a core diameter of $100~\mu m$, a clad/core ratio of 1.2,
and a NA of 0.22.
The fibers are inserted into a 3-layer PVC-kevlar tubing (from Northern Lights
Cable, U.S.A.) for protection and fitted with end connectors for mounting purposes.

At the output end, the bare fibers are glued onto small blocks,
called slitlets, that once assembled form the fiber slit.
At the input end the fibers are held by thin metal sheets,
called fieldlets, that once stacked on top of each other,
form the two-dimensional integral field at the backside of the lens array.

The fiber module contains 256 fibers, arranged in groups of 16.
The fiber arrays are self-contained elements that can be mounted and
de-mounted from end to end within the fiber module for the purpose of repair or replacement.
Additionally, it is also possible to remove the entire fiber module from the
PMAS instrument and replace it by an updated or modified version.
Both the lens array at the fiber input and the first spectrograph lens at
the fiber output are optically matched to the fibers, using optical gel
(code~0608 from Cargille Laboratory).
The effect of index-matching the fiber ends is twofold. Firstly,
reflection losses at glass to air interfaces are reduced and some
imperfections of the fiber end surfaces are cured by the gel.
Tests done by Schmoll et al.\ 2003 and by Kelz et al.\ 2003 demonstrate that fibers,
if index-matched to a glass plate or a lens array, show an increase
in flux transmission of approximately 10~\% , and a reduced fiber-to-fiber variation.

\begin{figure}[t!]
\vskip 2mm
\includegraphics[scale=0.32,angle=-90]{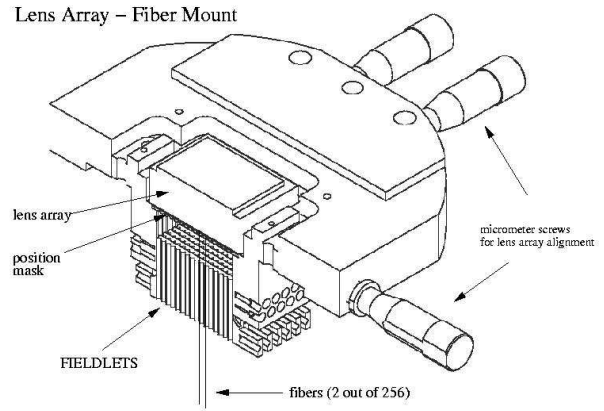}
\vskip 3mm
\caption{Cut view of the lens array mount, illustrating how the fieldlets with the
         fibers are positioned behind the adjustable, monolithic lens array.}
\label{fig:larrfblk}
\end{figure}

The current PMAS lens array of 16~$\times$~16 elements forms 256 micropupils
with diameters of 43 to 85~$\mu$m (see Table~3) simultaneously, which need to 
be imaged onto 100~$\mu$m fibers each. The common mount that holds the lens
array and the fiber ends ensures, that all fibers are placed at the
correct lateral and focal position with respect to the micropupils to within
$\pm 5~\mu m$ (see Fig.~\ref{fig:larrfblk}).

Instead of a permanent bound, the solution found for PMAS consists of a mount that
both holds the lens array and the fiber arrays mechanically.
A position mask is situated 0.2~mm behind the
backside of the lens array. The position mask consists of 256 holes
with an inner diameter of 150$\pm3$~$\mu$m and a pitch of 1$\pm0.003$ mm.
The mount provides a frame with grooves in a
1~mm spacing into which the fieldlets, carrying the fibers,
can be inserted and secured.
The fieldlets are manufactured in such a way that once they are
inserted in the mount, the fiber ends are located at the micropupil plane
(backside) of the lens array.
While the fiber assemblies and the mask are fixed in position,
it is possible to translate and rotate the lens array, using three micrometer
screws, for the purpose of alignment.

\begin{figure}[h!]
\plotone{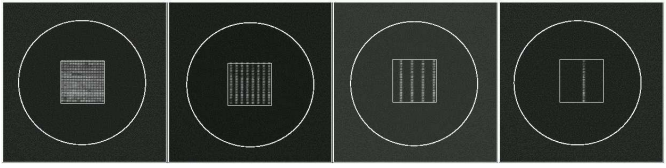}
\caption{Snapshots of the (backwards illuminated) PMAS lens array.
        Calibration mask stops can be inserted in front of the lens array
        to partly block the IFU for calibration purposes (see text).
        From left to right: the full FOV of 16 $\times$ 16 elements (white square),
        every 2nd column blocked, every 4th column open, only one column, e.g. every  16$^{th}$ spectrum illuminated.}
\label{fig:calmasks}
\end{figure}

The lens array mount can be equipped with one out of four calibration masks,
to cover vertical strips of the lens array (see Fig.~\ref{fig:calmasks}).
As a result, only every 2nd, 4th, or 16th fiber of the pseudo-slit is illuminated,
allowing one to measure the faint extended wings of the spectral profiles
in the cross-dispersion direction. The actual {\em measurement} of the profile
is a method of accurately modeling the spectra for optimal extraction
techniques, the correction of cross-talk, and to determine an
instrumental stray light model (see \S~\ref{sec:DRS}).

It is essential to mount fibers without induced stress, caused by
pressure, pull or strong bending, which otherwise increases the focal 
ratio degradation (FRD) and effects the performance.
To avoid any stress, the fibers are inserted in protective tubing ensuring
only moderate bending radii. Additionally, the fiber module
features a fiber-loop box, to avoid fiber 
breakage, and to provide a reservoir of extra fiber length which is
needed during assembly and for adjustments of the fore optics focal length.

\begin{figure}[h!]
\plottwo{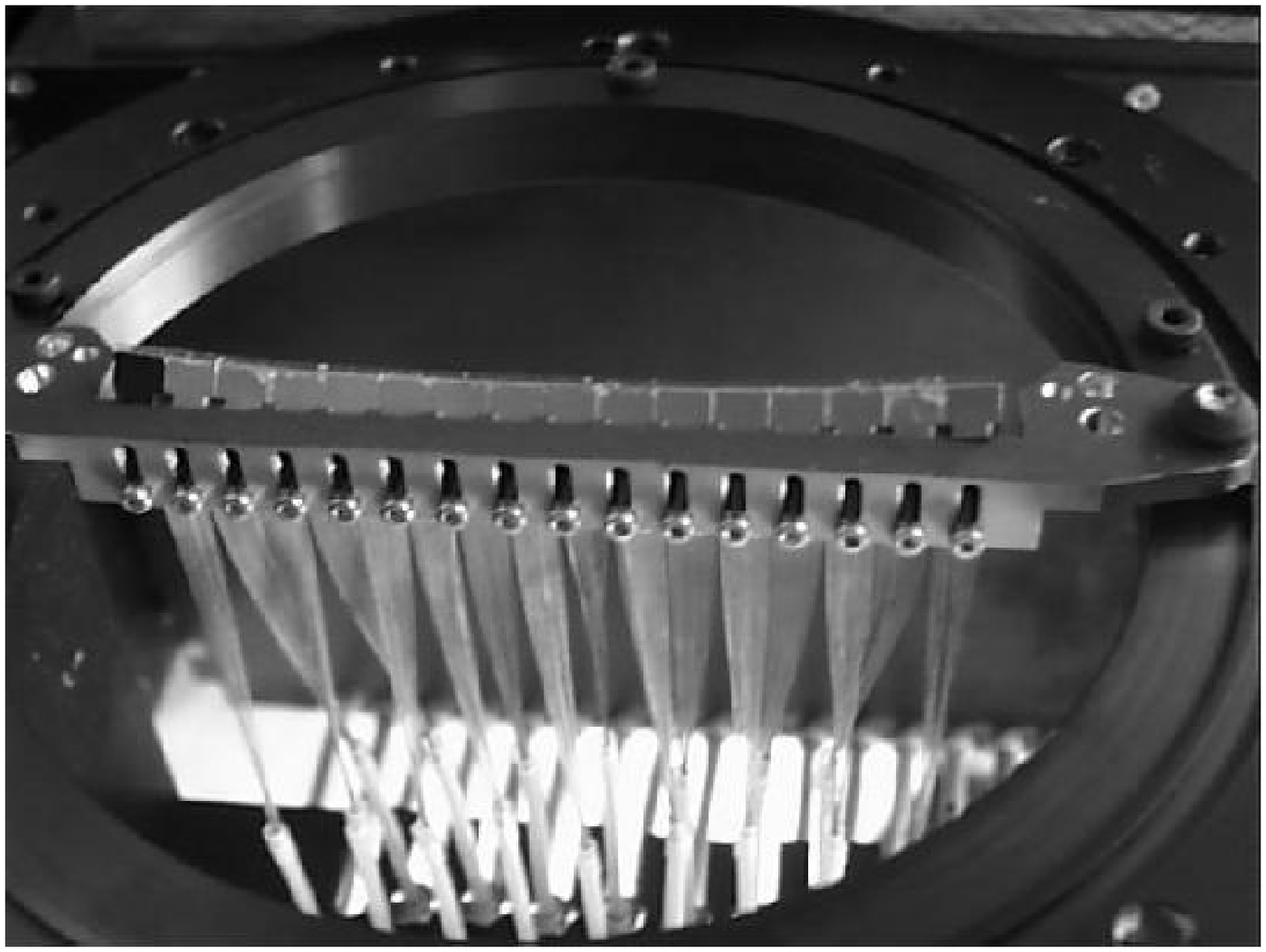}{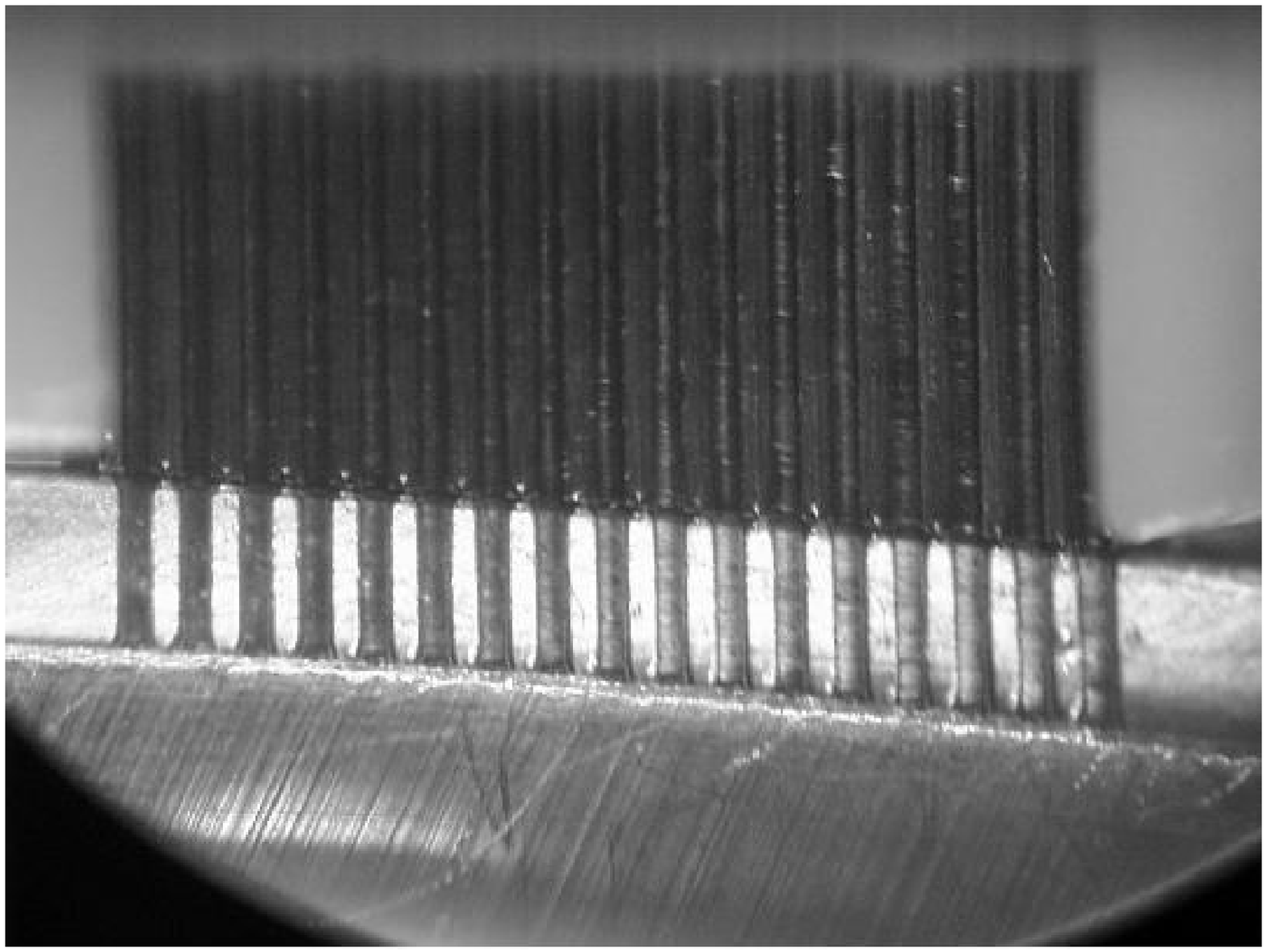}
\caption{Left: photo  of the assembled fiber-slit, which consists of 
16 blocks with 16 fibers each, yielding 256 fibers in total. Right: a 
detailed view of a single slitlet. Note the accurate position of the 
fiber end-faces, the overall curvature of the slit, and a 
17$^{th}$ spare fiber per block.}
\label{fig:fibslit}
\end{figure}

The output ends of the fibers are assembled onto 16 blocks (slitlets)
and arranged side-by-side to form an array of length 96 mm,
the fiber-slit (see Fig.~\ref{fig:fibslit}). The spacing from fiber
to fiber is 0.34~mm (=3.4 fiber core diameters), resulting in a  spectra-to-spectra distance,
which is sufficient to perform an interlacing nod-and-shuffle (beam-switching) mode
(Roth et al.\ 2004d, see also paper III).
The fiber-slit can accommodate double the number of fibers (512)
with half the spacing (corresponding to 7 pixel at the detector).
The spectrograph F/3 collimator accepts the whole fiber output cone
without the need of additional beam-converting microlenses.
The optical design of the spectrograph requires a curved fiber-slit, 
that directly couples to the surface of the first collimator lens, 
with optical gel matching the refractive indices.

\subsection{The Detector Subsystems}

PMAS is equipped with two cryogenic CCD systems: one for the spectrograph
camera, and another one for the acquisition and guiding (A\&G) camera. 
Except for some detector head and cable details, both subsystems 
share the same layout, based on ESO designs which were copied and adapted
with permission. The CCD subsystems each consist of a commercial 
IR Lab model ND-8 cryostat, detector head with detector head electronics, 
cables, and a CCD controller. The spectrograph detector head was built according
to the VLT design as described by Lizon (1997), whereas the A\&G camera
detector head was made
on the basis of an earlier design for La Silla, which was also used for the
MONICA instrument (Roth 1990). The CCD controllers are modified copies of
ACE (ESO Array Control Electronics), which was originally developed by
Reiss (1994) for the ESO VLT technical camera systems. ACE is based on a 
transputer--DSP architecture, employing the (now discontinued) INMOS T805/T225
transputer chips and a Motorola 56001 DSP. Both controllers are 
connected via transputer links to an on-board ULTRA SPARC station via SBUS 
interface cards. The SPARC is supervising the ACEs and handles CCD requests 
as well as the associated image data streams. 

The CCD software is a client-server based application written by TF.
The original INMOS server program was enhanced with an EPICS interface layer
(see \S~\ref{sec:INSTRUMENTCONTROL}), 
providing all necessary control functions such as initialization,
wipe, dark/\-bias/\-real exposures, window readout, binning, etc.\ from Unix
shell scripts. Shutter control is also performed from this environment.
A most useful feature of the CCD software is that each relevant ACE hardware
and software parameter is reflected (and updated) in the EPICS real-time
database, providing seamless access from high-level software and
advanced scripting without requiring any low-level software updates or
modifications. The PMAS nod-shuffle mode (described in paper III)
makes particular use of this scheme of low-level scripts and database 
parameters.

PMAS is currently equipped with 3 fully functional cryostat/detector head
subsystems, and 5 ACE controllers (3 of which are spare units). 
There is one cryostat/detector head for the A\&G camera system with a thinned 
1K$\times1$K SITe TK1024 chip, whose pixel size of 24$\mu$m is matched to 
the optical system to yield a plate scale of
0.2~arcsec/pixel. The A\&G camera field-of-view is 3.4$\times$3.4~arcmin$^2$.
The two remaining cryostats have alternative CCD configurations with (1) a
single SITe ST002A 2K$\times$4K chip, pixel size 15 $\mu$m square,
and (2) a mosaic of two such chips, covering the whole useful focal plane
of the spectrograph camera of 4K$\times$4K 15 $\mu$m pixels. Owing to the
external focus of the spectrograph optics, interchanging CCD cameras (1)
and (2) is a rather simple mechanical operation.

\subsection{Instrument Control}
\label{sec:INSTRUMENTCONTROL}

\subsubsection{Instrument Control Electronics}

Except for the exchange of gratings and filters, PMAS is operated under
remote control without direct human interaction at the instrument. The instrument
control electronics was designed as a self-contained subsystem and built
into a dual 19'' electronics rack. It is mounted in a rigid custom-designed 
aluminum frame, which in turn is suspended in the main frame.
It contains the following major components: a VMEbus computer as
master hardware controller, a SPARC workstation for data acquisition 
and high-level instrument control software, DC and stepping motor controllers,
motor power amplifiers, a shutter controller, a calibration unit with 6
shutter-controlled continuum and spectral line lamps, various power supplies,
and miscellaneous support and auxiliary electronic units. The substantial
amount of dissipated heat (typically 600~W) is removed through 8 front-panel
mounted fans when the instrument is operated off-telescope, or by means
of two liquid-cooled heat exchangers during observations at the telescope.
The VME subsystem contains various boards for dedicated control functions:
single board computer MVME167 (VME master), running under the real-time 
operating system VxWorks, PMAC motor controller 
for up to eight drives, accompanied by a 64-channel digital 
input/output board for general purpose I/O (sensing switches, controlling lamps
and shutters), MAC4-STP stepper motor controller for four stepping
motors, and a temperature monitor board for RTD sensors.

The following motorized functions are served by the electronics subsystem:
focusing the spectrograph collimator and camera,
grating rotator, focusing the A\&G camera optical system, 
filter exchange stage of the A\&G Camera, linear stage for the
exchange of fore optics collimator lenses (changing the IFU scale/magnification),
linear stage to adjust fore optics for the difference in focal length of
different collimators, fore optics exchange mechanism for different order 
separating filters, linear stage to deploy/retract the calibration unit
into/from the telescope beam, linear stage to insert the relay optics of
the A\&G camera. In addition to these motor-controlled functions, there
are 6 individually controlled mechanical shutters for the lamp modules
of the calibration unit, the fore optics shutter, and the spectrograph slit 
shutter which resides inside of the spectrograph collimator. 
The linear stage motors are protected against
malfunction through a variety of redundant inner and outer limit switch pairs,
as well as fail-safe interlock schemes for those drives which are subject
to potential collision states. All motors are equipped with gears, precision
linear or circular position encoders, and tacho encoders. The grating rotator,
which yields a positioning accuracy of 5~arcsec, is equipped
with a DC motor, coupled to a HarmonicDrive gear and followed by a precision 
rotator stage, both of which are designed to exhibit zero backlash.

The VMEbus computer, the SPARC workstation, and an associated terminal server 
are connected via Ethernet and a layer-3-switch to the local area network (LAN).
The terminal server offers console access to each computer and the PMAC
motor controller, which is allowing to startup the instrument and boot the
computers from remote login, as well as providing direct access for
diagnostics purposes.

\subsubsection{Instrument Control Software}

\begin{figure}[t!]
\plotone{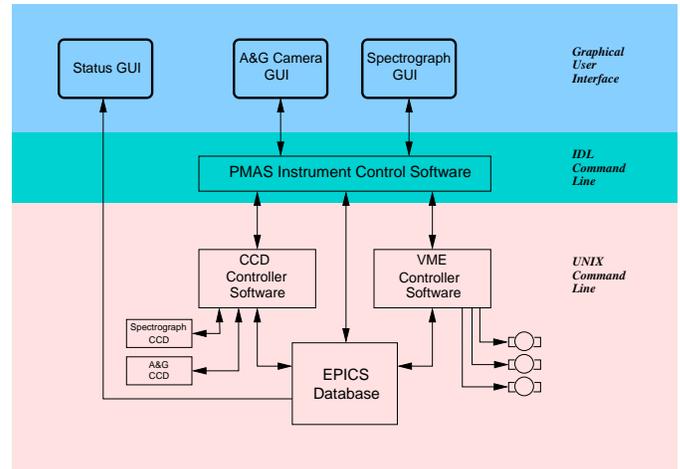}
\caption{PMAS instrument control software.}
\label{PMAS_ICS}
\end{figure}

The PMAS instrument control software has a hierarchical 
structure (Fig.~\ref{PMAS_ICS}), 
which is implemented on two main computers (target/host) and a
number of embedded controllers. The host system is a 
Unix workstation, running EPICS, the ``Experimental and Industrial 
Control System'' \footnote{http://www.aps.anl.gov/epics}.
The target system is a VMEbus single board computer, running under the 
real-time operating system VxWorks. The Input/Output Controller part of 
EPICS is implemented on the target.

The low-level target software consists of a state program, the EPICS real-time database
accompanied by hardware drivers and PLC and motion programs
running on the PMAC motor controller. The EPICS database is a graphically
designed software entity, which, once booted, runs permanently, similar to
a common hardware device. Instrument control and logging functions are triggered 
periodically, or driven by events or interrupts. A useful feature
of the EPICS real-time database is the availability of each instrumental
parameter known to the the instrument control software. 
Host and Input/Output Controllers (IOC) communicate 
via LAN. EPICS offers a communication mechanism called {\em Channel Access},
providing transparent network access from host and IOCs to  
EPICS database process variables just by name. 

The low level software is broken down into CCD-related tasks (CCD Controller)
and other hardware-related tasks (VME Controller), both of which are coordinated by
and whose status are reflected in the EPICS database. From the
user point-of-view, all low level tasks are accessible from a
Unix command line interface with a variety of low level shell scripts.
In addition, a generic EPICS graphical user interface provides status
information about all major mechanical devices, lamps, and other parameters like
temperatures, CCD cryostat evaporating N$_2$ flow rates, etc.\ 
(PMAS Status Window).

A second level command line interface is provided through a collection of 
IDL scripts, interfacing to the low level software and EPICS database, and
shielding the user from the detailed knowledge of device parameters. 
All basic instrument control functions like starting CCD exposures, switching 
lamps on/off, setting the grating, focus, and so forth are accessible through these 
comprehensive scripts, which also include basic help features, as well as logging 
and diagnostic functions.
Elementary scripts can be easily combined to form more complex macros for complete
measurement cycles including the evaluation and visualization of results.

On top of the PICS command line interface, two major GUIs (also written in IDL)
are available for observing at the telescope: the spectrograph CCD control
interface, and a corresponding tool for the A\&G camera.

\subsection{Data Reduction Software}
\label{sec:DRS}

Based on experience gained with the MPFS instrument at the 6m BTA in
Selentchuk (Si'l\-chen\-ko \& Afanasiev 2000), a software package, 
developed by Becker (2002), is used for quick-look inspection of data quality 
at the telescope ({\em P3d\_online}) or final data reduction ({\em P3d}). 
The code is written in IDL and comprises more than 1200 routines for 
processing the raw data and subsequent steps of data analysis, e.g.\ 
subtracting bias and dark frames, CCD pixel-to-pixel response variation,
removal of cosmic rays, tracing, flexure compensation, swath extraction or 
profile-fitting extraction, straylight modeling, wavelength calibration, 
and wavelength-dependent fiber response calibration. There are 
various tools for the visualization of stacked spectra,
maps, and individual (or coadded) spectra, for aperture spectrophotometry,
datacube PSF-fitting routines, atmospheric dispersion compensation, and
so forth (programs {\em monolook}, {\em cube\_viewer}, and many others). 
Some elements of these tools were found to be useful prototypes for the 
development of the {\em E3D} visualization tool which is commonly available as open
source code through the Euro3D consortium (S\'anchez 2004, S\'anchez et 
al.\ 2004). The {\em P3d} and {\em P3d\_online} pipelines are accessible through
GUIs, facilitating the access of science data and calibration files,
and the setup or modification of parameters. It is also possible to
operate the pipeline from the IDL command line, using scripts instead
of the GUI. P3d has also been successfully applied to data other than
from PMAS, e.g.\ for MPFS, SPIRAL, INTEGRAL, and the VIMOS-IFU. More
recently, {\em P3d\_online} was modified to accommodate and visualize
data obtained with the new PMAS fiber bundle IFU 
({\em PPak\_online}, see paper II).

\section{Instrumental Performance}
\label{sec:PERFORM}

\subsection{Assembly, Integration, and Tests}

During the development phase, the various sub-com\-pon\-ents of PMAS
were tested in the laboratory facilities of the AIP. These include tests in
the mechanical, optical and electronics labs, as well as system
checks at the AIP telescope simulator.
The spectrograph optics were tested at the facilities of Carl Zeiss Jena 
(Roth et al.\ 2000b), see \S~\ref{sec:FSPECTEST}. The image quality of all
256 microlenses of the lens array was measured at the AIP,
see \S~\ref{sec:LARRTEST}.
The complete fore optics, including the internal calibration unit were
pre-assembled and tested at the optical laboratory. Extensive fiber tests,
including throughput, focal ratio degradation and stress behaviour
were undertaken using a specialized fiber testbench (Schmoll 2001, Schmoll et
al.\ 2003). The three CCD cameras (A\&G, single chip spectrograph, mosaic spectrograph)
were assembled in the detector and clean room facilities and optimized with
respect to dark current, readout noise, bias, gain, etc.\ using the
AIP photometric testbench (Fechner et al.\ 2000).
A commercial interferometer was used to assess the optical quality  
of the folding mirrors,
the backside of the lens array, and the dewar windows.

\begin{figure}[t!]
\vskip 3mm
\plotone{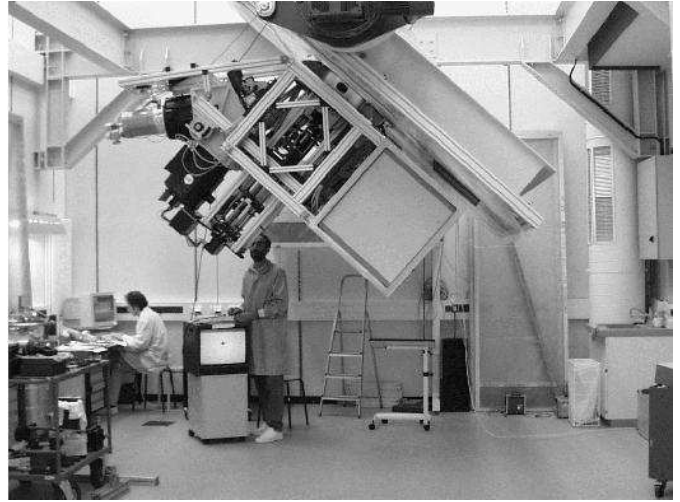}
\caption{The PMAS instrument during system performance tests at the 
AIP telescope simulator. A star simulator can be mounted on top of the device.}
\label{fig:telsim}
\end{figure}

During the integration of the instrument, PMAS was mostly mounted to the telescope
simulator at AIP  (Fig.~\ref{fig:telsim}). This allowed good
access to the instrument, helped in the alignment of the long fore optics
tray, and, importantly, provided an option to tip-tilt the instrument.
Measurements, often with high spatial magnification, were taken at various
instrument inclinations, to evaluate the the mechanical and thermal
stability (see \S~\ref{sec:TELMODSTAB}, \ref{sec:FSPECSTAB}), and
\S~\ref{sec:THERMALSTAB}, respectively).
Using a setup with pinholes and relay lenses, an artifical star was simulated
and imaged onto the IFU and the A\&G system.
The actual spectrograph performance (i.e. PSF, dispersion,
coverage, relative grating response, focus behaviour, etc.) was measured
repeatedly. These measurements were also used to evaluate and optimize the
fiber module, i.e.\ the mounting of the fiber slit, the quality of the index-matching,
the alignment of the lens array, the fiber-to-fiber response, etc.
\vskip 10mm

\subsection{Spectrograph Optics}
\label{sec:FSPECTEST}

The spectrograph optical system, consisting of the collimator
and camera subsystems, was delivered by the manufacturer with an
end-to-end acceptance test, which was performed in collaboration with
AIP. Details of this test were reported in Roth et al.~2000b, and here
we shall only summarize some key features of the setup and the major results. 
In the absence of the PMAS spectrograph CCD camera, which was still under
development at the time, a commercial cryogenic CCD camera
(Photometrics AT200/CE200A), equipped with a blue-sensitive, backside-illuminated
TK1024 CCD, was used to observe the image of an artificial star in
the focal plane of the camera. The spot was enlarged with an
$\approx$20$\times$ magnifying microscope objective, yielding a scale of 
1.22$\pm$0.01~$\mu$m/pixel. Collimator and camera were linearly
aligned and illuminated so as to create a spot of controlled size
at the collimator input, and uniform illumination at the nominal
focal ratio of f/3 in the pupil plane. The results of the image
quality tests are listed in Tab.~\ref{tab:FSPECSPOTS} in terms of
diameter of 80\% encircled energy.

The overall transmission was not measured end-to-end, but estimated
from the acceptance test records of the individual lens coatings, 
and from tabulated throughput data for the glasses and CaF$_2$. 
The total transmission over the wavelength range 450--1000~nm 
is $\approx$80\% and almost constant. Towards the blue, it drops 
nearly linearly to 70\% and 55\% at 400~nm and 350~nm, respectively.

A differential flexure test at the premises of Zeiss, which was performed
by tilting the collimator-camera system with detector on 
a rigid platform in steps of 15$^\circ$ from 0 to 90$^\circ$,
yielded an image shift of 7.5$\pm$1$\mu$m over the whole tilt range, which was
reproducible and gave no visible indication of PSF-variation.
Note that during this test, the linear stages of the focusing
lenses (both collimator and camera) were {\em clamped on both sides}, thus
eliminating additional degrees of freedom. It was discovered later
during the PMAS assembly, integration, and test phase that
a lack of stability of those mechanisms is responsible for the
noticeable flexure of the integrated instrument (see \S~\ref{sec:FSPECSTAB}).

\pagebreak
\centerline{Table~4}
\vskip 2mm
\label{tab:FSPECSPOTS}
\begin{center}
\begin{tabular}{crcc}
$\lambda$~[nm] & D80 design & D80+pinhole & D80 measured \\
\hline\hline
       &              &              &              \\
\multicolumn{4}{c}{0$^\circ$} \\ \hline
365    &  18.5        &  24.5        &  22.5        \\
436    &  11.2        &  17.2        &  17.5        \\
546    &   8.5        &  14.5        &  17.5        \\
852    &  12.0        &  18.5        &  27.0        \\
       &              &              &              \\
\multicolumn{4}{c}{3.5$^\circ$} \\ \hline
365    &   9.7        &  15.7        &  43        \\ 
436    &   5.5        &  11.5        &  27        \\ 
546    &   8.5        &  14.5        &  24        \\ 
852    &   6.3        &  12.3        &    -       \\ 
       &              &              &            \\
\multicolumn{4}{c}{4.9$^\circ$} \\ \hline
365    &  12.5        &  18.5        &  33        \\ 
436    &  22.9        &  28.9        &  25        \\ 
546    &  11.1        &  16.1        &  22        \\ 
852    &  10.5        &  16.5        &    -       \\ 
       &              &              &            \\
\multicolumn{4}{c}{6.0$^\circ$} \\ \hline
365    &  19.0        &  25.0        &  32        \\ 
436    &  31.1        &  37.1        &  27        \\ 
546    &  20.3        &  26.3        &  26        \\ 
852    &  19.0        &  25.0        &    -  \\ 
\end{tabular}
\end{center}
{\footnotesize{\em Encircled energy as function of
field angle and wavelength. Column 1: wavelength in nm, 
col.~2: predicted D80 according to optical design,
col.~3: predicted D80, convolved with 10$\mu$m pinhole, col.~4: measured D80.
(D80: diameter of 80\% energy concentration [$\mu$m])}}

\subsection{Lens Array}
\label{sec:LARRTEST}

\begin{figure}
\plotone{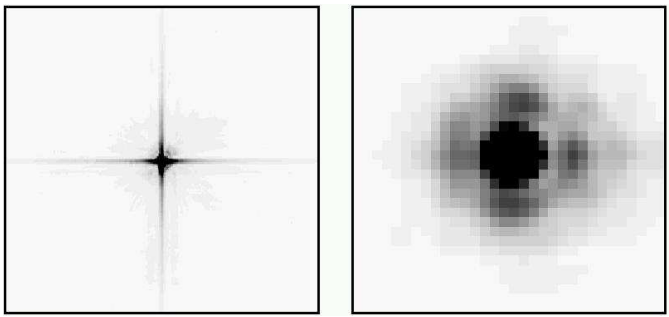}
\caption{Lens array PSF, obtained with cryogenic CCD over high dynamic
range. Left: single lenslet PSF, average from 101 single CCD exposures,
greyscale stretch at 1\% peak intensity;
frame size: 400$\times$400 pixel $\approx$~0.5$\times$0.5~mm$^2$.
Right: PSF core detail with Airy pattern, average over all 256 lenslets 
of the whole 16$\times$16 lens array. 
  \label{fig:DEEPSPOT}}
\end{figure}

The lens array image quality was assessed in lab tests using
the same cryogenic camera as described in \S~\ref{sec:FSPECTEST}.
A 10$\mu$m pinhole was projected to infinity with a f=160mm
achromatic lens, illuminating the lens array element under study.
In its focal plane a 21-fold demagnified image of the pinhole was 
observed with the 20$\times$ microscope objective and the CCD camera.
Due to the demagnification, the diameter of the pinhole 
image is negligible with respect to the Airy pattern of the spot
such that the observed figure can be interpreted directly as the 
PSF of the lenslet.  The left frame in Fig.~\ref{fig:DEEPSPOT} 
shows the superposition of 101 individual exposures of a single spot, 
which exhibits a very good S/N-ratio even at large radii from the 
centroid of the spot. 
The PSF core has a FWHM of 4.8$\mu$m. The first Airy diffraction 
minimum occurs at a radius of $\approx$~4.9~$\mu$m. There are significant 
contributions of diffraction, aberration, and stray light to the extended 
wings of the PSF at low intensity levels.

\begin{figure}[h!]
\plotone{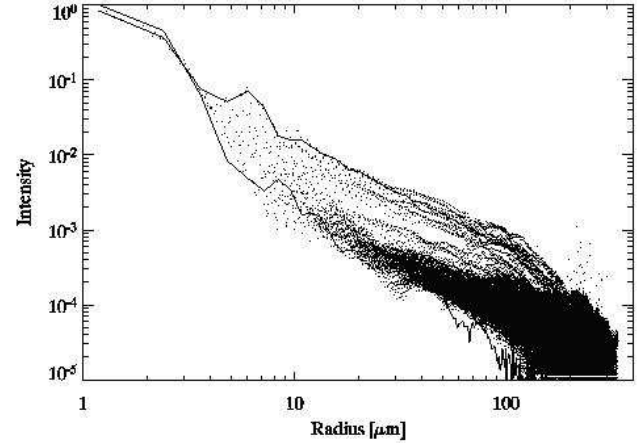}
\caption{Radial plot with normalized intensity of each pixel in 
  Fig~\ref{fig:DEEPSPOT} as a function of distance from the
  center. The upper full drawn curve follows the intensity along
  the vertical diffraction spike, the lower curve along a diagonal
  line with 45$^\circ$ inclination.
  \label{fig:LARR_RADIALPLOT}}
\end{figure}

\begin{figure}[h!]
\plotone{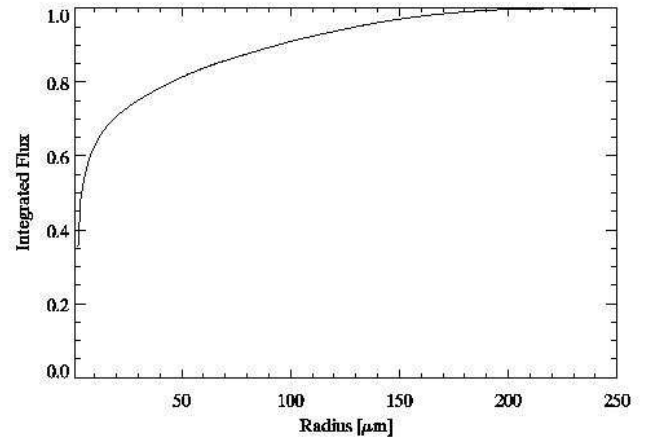}
\caption{Encircled energy for the spot shown in Fig~\ref{fig:DEEPSPOT}. 
  \label{fig:LARR_GCURVE}}
\end{figure}

Fig.~\ref{fig:LARR_RADIALPLOT} shows a radial intensity
plot of the spot. The dispersion over different azimuthal positions is 
due to the difference between the enhanced intensity along the pronounced 
diffraction spikes and the lower intensity in between.
Fig.~\ref{fig:LARR_GCURVE} shows a normalized curve-of-growth for the spot,
plotting the encircled flux versus radius. 
The extended parts of the PSF, i.e.\ diffraction spikes and stray light,  
make a significant contribution to the total flux. 80\% energy
concentration is reached at a radius of $\approx$45$\mu$m. By convolving the
nominal 43$\mu$m micropupil (0.5~arcsec/spaxel magnification) with
this PSF and measuring the flux within the 100~$\mu$m aperture of
an ideally centered fiber, we find a typical light loss of 25-30~\%.
We have measured 80\% energy concentration values for all lenslets
of the first delivered array, as well as for random selections of
lenslets of two 16$\times$16 element arrays, and of one 32$\times$32 
element array from a second batch. 
We determined the following spot concentration statistics for these 
four devices, respectively: 
({\em 40}--55--{\em 65}), ({\em 30}--49--{\em 72}), ({\em 30}--57--{\em 70}), 
({\em 30}--51--{\em 62}), given as 10\% -- median -- 90\% percentiles for
radii of 80\% encircled energy [$\mu$m].
We conclude that the performance is comparable to the ``epoxy on glass''
type of lens array in the study of Lee et al.~2001, but a factor
of two better than the ``crossed cylindrical lens'' type made from
fused silica, for which a device was tested which has similar parameters
as the PMAS lens array (1mm pitch, f/5.5). Note that our results re\-present
upper limits since the raw frames were not corrected for a possible
additional stray light contribution from the microscope
objective as advised by Lee et al.\ 2001.

\subsection{Instrumental Throughput}

\begin{figure}
\plotone{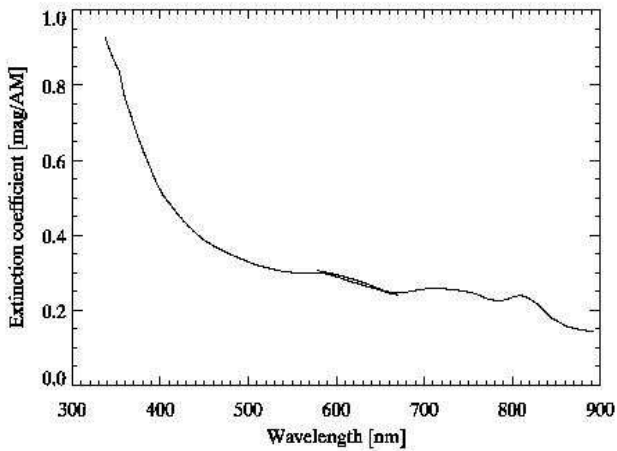}
\caption{Atmospheric extinction at Calar Alto on Aug.\ 1, 2003, determined
 from standard star measurements. The observations cover airmasses from
 1 to 2 at two wavelength settings (note the overlap at 600~nm).
 The Calar Alto extinction monitor recorded values between 0.32 to 0.4 mag in V.}
\label{fig:extinction}
\end{figure}

The instrumental throughput was measured by observing
spectrophotometric standard stars and correcting for the atmospheric
transmission and reflection losses of the telescope.
The observation of flux standards as routinely performed during most 
observing runs has shown that the atmospheric conditions at 
Calar Alto are often quite variable, making it difficult to establish
a reliable extinction (see also Hopp\&Fernandez 2002).
On August 1, 2003, however, the flux standard  BD+33d2642 (V=10.83,
Oke 1990) was observed repeatedly under photometric conditions over the airmass
range of 1.0 to 1.95, resulting in an accurate determination of 
the extinction curve for this night (Fig.~\ref{fig:extinction}).

The ratio of observed to expected photons defines the overall efficiency $\eta$,
which includes the instrument, the atmosphere and the telescope:

\begin{equation}
\eta = \frac{\#phot_{obs}(\lambda)}{\#phot_{exp}(\lambda)}
     = \eta_{atm}~\eta_{tel}~\eta_{instr}
\end{equation}

The atmospheric extinction coefficients for each wavelength were calculated
in the standard way, using the repeated observations of the star at
different airmasses (see Fig~\ref{fig:extinction}), which 
yielded $k_{ext}$=0.3 mag in V and $\eta_{atm}$=0.76.

The reflectivity of the primary mirror is measured routinely at Calar Alto.
Assuming a similar value for the secondary reflectivity, the telescope
efficiency was estimated to be $\eta_{tel}$=0.54 in V at the time of
observation.
This allowed the determination of the PMAS instrumental response for the setup using
the lens array IFU with $8'' \times 8''$ FOV and the V300 grating.
Instrumental throughput values for other configurations were obtained by
bootstrapping from this measurement to the relative flux responses of domeflat
exposures, using other gratings that are available for PMAS.

\begin{figure}[t!]
\plotone{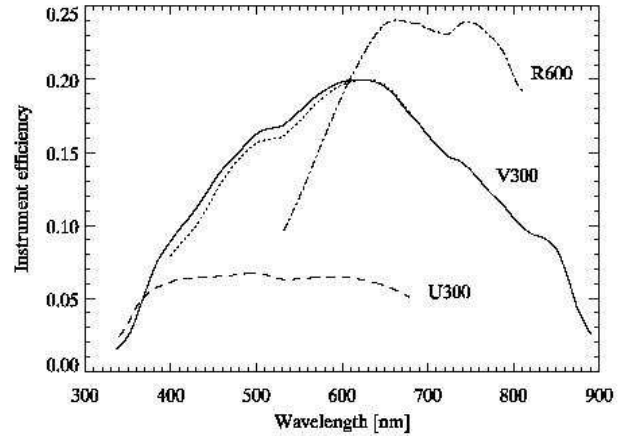}
\caption{Plots of PMAS instrumental efficiencies, indicating the different response
of a grating blazed at V (solid curve), U (dashed curve) and R (dashed-dotted curve). The efficiency of the V1200 grating (dotted curve) is similar to the V300.}
\label{fig:efficiency}
\end{figure}

Figure~\ref{fig:efficiency} plots the pure 
PMAS instrumental efficiency $\eta_{instr}$, i.e.\ the throughput from
the telescope focal plane to the detector. The maximum efficiency is
found to be near 20\% at $\sim$600~nm for gratings blazed at V.
Redwards of 600~nm, gratings blazed at R show a peak efficiency of 24\%. 
Gratings blazed at U are superior only at wavelengths below 360~nm. 
While the throughput depends on the blaze function of the gratings, 
it is almost independent of the groove density (e.g. 300, 600 or 1200 l/mm).

\subsection{Fiber-to-Fiber Throughput Variations}

\begin{figure}
\plotone{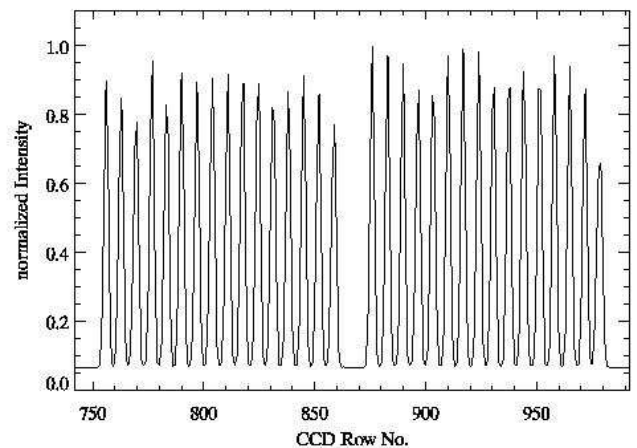}
\caption{Fiber-to-Fiber throughput variation for two typical groups
of 16 spectra (bias level not subtracted).}
\label{fig:FIB2FIB}
\end{figure}

Due to the combined effect of less than perfect image quality for
individual lens array lenslets, and  alignment, coupling, and the
transmission details of individual fibers, there are non-negligible
throughput variations from spectrum to spectrum. Fig.~\ref{fig:FIB2FIB}
gives an impression for two groups of 16 spectra, corresponding to
slitlets no.\ 7 and 8 near the center of the FOV.

Formally, the throughput variation measured from the flux obtained
from extracted flatfield spectra amounts to 12\% r.m.s. for all 256
spectra, or to 6\% r.m.s. if three outlier groups are excluded. 
This method of
characterization, however, is not a fair measure since the statistical 
distribution of fiber transmission values is non-normal, and the systematic
effect of vignetting in FSPEC is not taken into account. We rather
note that 77\% of all spectra are performing with $>$80\% throughput
(T) in terms of peak transmission, 
17\% with 70\%$<$T$<$80\%, and 6\% outliers with T$<$70\%.

\subsection{Spectrophotometric Accuracy}

\begin{figure}[!h]
\plotone{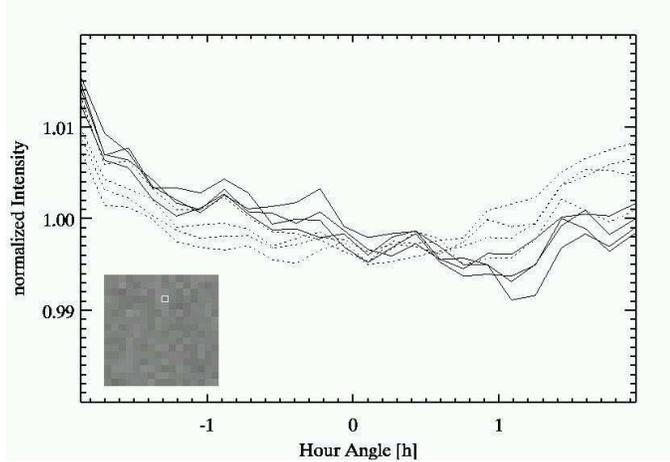}
\caption{Continuum flatfield intensity variation of fiber \#~200, 
telescope tracking over 4 hours. Insert: flatfield map, spaxel \#~200
indicated.}
\label{fig:fiber-var}
\end{figure}

The instrumental accuracy of spectrophotometric measurements is 
strongly dependent on the reliability of flux calibrations, i.e.\
the stability between measurements of a science target, and of the
corresponding flux standard(s). Owing to variable atmospheric conditions
at the site, an end-to-end validation seemed to be a less than trivial
undertaking. Instead, internal spectrophotometric stability tests
were taken in daytime or during nights with poor weather conditions,
allowing the telescope to track on an imaginary target for many
hours, and taking internal continuum and spectral line lamp exposures
with the V300 grating over regular time intervals. Fig.~\ref{fig:fiber-var} 
illustrates the result from a series of 24 such measurements at a declination of 
37$^\circ$~13$^m$ over an hour angle range of -2$\ldots$+2 hours.
The continuum flatfield frames were bias-subtracted, traced,
extracted, and wavelength-calibrated in the usual way (\S~\ref{sec:DRS}).
The inspection of maps at selected wavelengths (or coadded over wavelength
intervals), normalized to the frame taken at HA=0, yields varying patterns
which are different from unity, indicating that the flatfielding process
was less than perfect. As an example, Fig.~\ref{fig:fiber-var}
shows a plot of this variation vs.\ hour angle for the (arbitrarily chosen)
fiber no.~200. The eight curves were obtained by averaging over 100 spectral
pixels each, in increments of 100 pixels from a starting wavelength of
$\approx$360nm. The blue bins are plotted as full lines, while the red bins 
are plotted as dotted lines for clarity. The reddening trend towards larger
hour angles is seen for all fibers, indicating perhaps a color temperature
change of the lamp and not an intrinsic feature of the IFU + fiber bundle.
Nonetheless, a systematic variation in any wavelength bin, typically on 
the order of 1\%, is observed over the 4~h duration of this test. We interpret
this variation as the typical accuracy limit for long-term flux calibrations (albeit
a caveat regarding the unkown stability of the calibration lamp). As
the most likely cause for this behaviour, we adopt the hypothesis of
a subtle lever effect on the fiber fieldlets (Fig.~\ref{fig:larrfblk})
under a varying gravity vector, leading to small micropupil--fiber 
displacements and a subsequent change of the lens array--fiber coupling
efficiency. This hypothesis is supported also from a known
sensitivity during the (difficult) lens array--fiber alignment process.

\subsection{Mechanical stability of the A\&G unit}
\label{sec:stabAG}

To evaluate possible flexure effects within the acquisition and guiding
(A\&G) system, which consists of folding mirrors and focusing lenses
(\S~\ref{sec:TELMOD}), the image motion of an artificial star
in the focal plane of the A\&G camera was observed as a function of
orientation. During tests at AIP, the telescope simulator was moved to 
a variety of inclinations ranging from -30 to +45 degrees.
During commissioning at the CAHA 3.5~m Telescope, flexure measurements with
zenith distances ranging from 0 to 63 degrees were performed.
In both cases the average image shift was found to be $\sim$0.1 pixel
(= 0.03 arcsec) and $\sim$0.3 pixel (=0.06 arcsec) in the x and y direction,
respectively. This means that
residual image shifts due to flexure are an order of magnitude smaller
than the seeing disk and thus negligible for accurate guiding and offset
applications.

\subsection{Mechanical stability of the telescope module}
\label{sec:TELMODSTAB}

\begin{figure}[h!]
\plotone{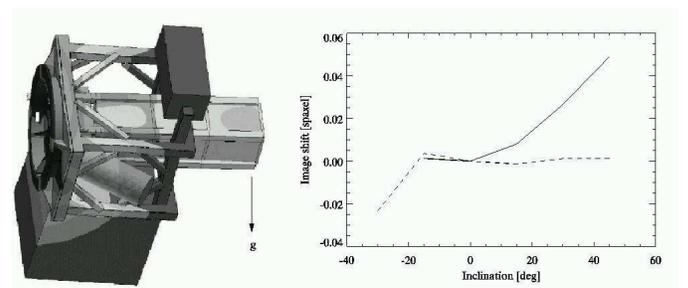}
\caption{Left: telescope module flexure, modeled with ANSYS FEA. 
The predicted maximum deflection at the end of the telescope module tower 
at an inclination of 90$^\circ$ is 0.1~mm.
Right: the measured image shifts in x (solid curve) and y (dashed curve) 
versus inclination are within a 1/10th of a lens array spaxel.}
\label{fig:TELMOD_FE}
\end{figure}

The telescope module contains a bench of $\approx$2m length which carries the
fore optics and the lens array IFU.
To estimate the flexure of the relative long optical bench, a finite element
analysis (FEA) was performed. In the resulting optimized design, 
the telescope module tower is
mounted only at its top end to the flange, but is not connected to the main frame.
The bench structure together with cover plates was estimated to
exhibit flexure less than 0.1~mm at 90$^\circ$ inclination of 
the instrument (Fig.~\ref{fig:TELMOD_FE}).

During flexure tests the AIP, using a star simulator with a variety of 
pinholes, the image of an artificial star was projected onto the lens array.
Spectrograph CCD exposures were taken for various inclinations of 
the telescope simulator, ranging between $-30^\circ$ and $+45^\circ$. The 
image of the star at the IFU was reconstructed the centroid  determined. 
The image shift on the IFU was found to be within $\sim$0.1~mm or $\frac{1}{10}$
of a spatial element (spaxel), which is in good agreement with the FE-model. 
Depending on the fore optics magnification this corresponds to an image 
motion of $0.''05$ to $0.''10$, i.e.\ well below the seeing FWHM.

An analysis of real exposures of point sources observed for half an hour
at the telescope, revealed no elongation of the images or asymmetries of
the PSF, which might have been caused by flexure of the fore optics.

\subsection{Mechanical stability of the spectrograph}
\label{sec:FSPECSTAB}

\begin{figure}[h!]
\plotone{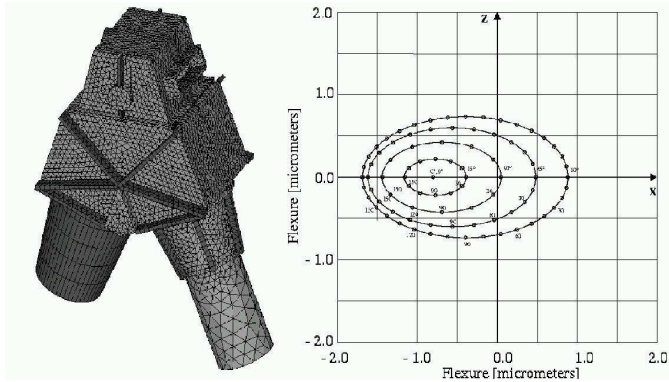}
\caption{Left: using an ANSYS FE-analysis, the stiffness of the spectrograph 
module was modeled. Right: the calculated stiffness of the spectrograph housing 
was expected to be well within 3~$\mu$m at all orientations.}
\label{fig:SPECMOD_FE}
\end{figure}

The mechanical design of the spectrograph mechanical structure was optimized
also with the help of ANSYS FEA (Dionies 1998). The final design
predicts an image motion of less than 3~$\mu$m at any telescope
inclination (Fig.~\ref{fig:SPECMOD_FE}).
Flexure tests undertaken at the AIP telescope simulator revealed
considerably larger image shifts than expected, which were found to be caused
by the rather instable focus mechanism of the spectrograph 
collimator and camera lenses,
respectively. Since the focus stages are integral parts of the optical
subsystems, it was not easily possible and considered a risk to replace the
poorly performing devices at the time of commissioning. After some
improvement by stiffening the
external mechanical interface to the focus stages, the remaining flexure had
to be tolerated for the time being.

\vskip 5mm
\vskip 5mm

\begin{figure}[t!]
\vskip -3mm
\centering\includegraphics[width=1.0\linewidth,clip]{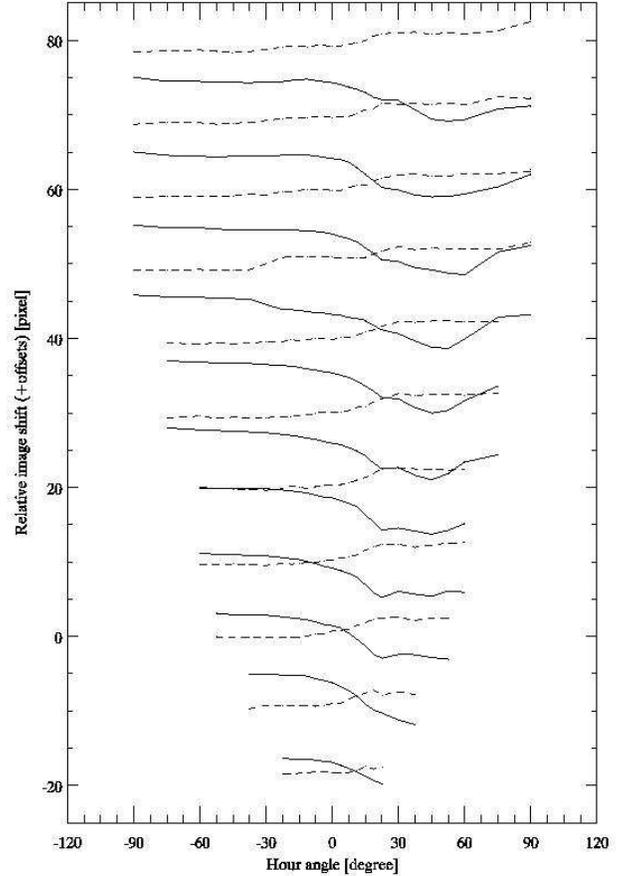}
\figcaption{Relative image shift
 in the spectrograph dispersion (dashed curves) and cross-dispersion (solid
 curves) direction versus hour angle (H.A.).
 The telescope was moved from East to West at eleven declinations (ranging from
 -20 to +80 deg). For better clarity, offsets were added for each declination strip.
 The stability is worse when pointing west of the meridian.
 \label{fig:flex_ha}
 }
\end{figure}

During some bad weather nights, flexure tests at the telescope were performed.
The telescope was pointed to various (up to 256) positions in hour angle (H.A.)
and declination (Dec.). At each pointing a short arc lamp exposure was taken with
the spectrograph CCD, a 1$\times$1 binned window around an isolated emission spot
was read out, and the resulting image shift computed.
Fig.~\ref{fig:flex_ha} plots the relative shift of the
centroid with respect to its position at zenith (H.A.=0:00 hr and $\delta$=+37.1 deg).

The image shifts in the spectral and the spatial direction differ.
Note that the shifts are relatively small for negative hour angles
(east of the meridian; rising objects). A significant shift seems to
occur during the transit period and for certain positive hour angles
(west of the meridian; setting objects).
In order to minimize the effects of flexure, resulting in a
degradation of spectral resolution and in an increase of
cross-talk between adjacent spectra, the critical positions near
transit must be avoided during long exposures, or the exposure time 
must be segmented accordingly.

\subsection{Thermal Stability}
\label{sec:THERMALSTAB}

The variation of spectrograph best focus as a function of temperature was
investigated at the telescope when an extended period of observing nights
with identical setup parameters during the nights November 10 --- 21, 2004,
presented us with the opportunity to evaluate focus series measurements
taken under similar conditions, but at different temperatures. 
The spot size was measured from the $\sigma$ provided by Gaussian 
fits to the emission line spots in spectral line lamp exposures in the 
dispersion (X) and spatial (Y) directions, respectively. 
From previous experience with gratings in 1st order it was known
that the two linear stages mounted in the collimator and camera 
lens barrels show a typical sensitivity of 150~$\mu$m travel from which 
on a significant departure from best focus sets in (10~\% increase of FWHM).
The focusing lens positions on their linear stages are motor-controlled with
a re\-so\-lution of 1.45~$\mu$m per encoder step. The setup for our
test was a factor of 3 more sensitive to displacement with respect to FWHM
in the dispersion direction since the J1200 grating in 2nd order was used
for medium spectral resolution observations, which benefits from anamorphic
demagnification. From the four nights of November 16--19, we were able to 
compare focus series measurements over a temperature range of
-2.6, +0.2, +2.4, +4.0~$^\circ$~C, respectively. The temperature value was taken from
the data file FITS header entry for the dome temperature. A typical
focus series is evaluated by plotting families of FWHM in X and Y vs.\
camera focus position, labeled by collimator focus position. Depending
on a criterion (minimal FWHM in X, in Y, or optimal encircled energy
for both directions), the curve presenting the absolute minimum over all pairs of
collimator and camera positions is selected, and the minimum parameters are chosen for
further use during observations. As a rule of thumb, the camera focus is
roughly a factor of 3 more sensitive to displacement as collimator. For the
analysis of thermal variation of best focus, we inspected the focus series
plots for shifts of the minima as a function of temperature. As a result,
we find a linear trend at temperatures 0...4~$^\circ$~C, and a steeper
gradient towards negative temperatures. Linear regression yields 
a camera focus shift of +19.9$\pm$0.7~$\mu$m/$^\circ$~ at positive temperatures.
Below 0~$^\circ$~C, we estimate a rate roughly 3$\times$ larger than this value
(however, based only on a single data point). We did no recognize a significant
temperature dependence of collimator focus over the observed temperature range,
particularly.

\subsection{CCD performance}

The 1K$\times$1K TK1024 of the A\%G CCD system has no major cosmetic
flaws and is operated with a conversion factor of k=1.4~${e^{-}}$/ADU and
a read noise of 4.6$e{^{-}}$. The quantum efficiency of the
chip as specified by the manufacturer is 65\% at 400nm, 80\% at 650nm, 
and 40\% at 900nm. 

The single chip spectrograph CCD system (1) with its standard
gain setting has a read noise of 4.3~e${^{-}}$ and a conversion factor
of 1.4~~e${^{-}}$/ADU. This CCD is read out over one functional amplifier.
It has an approximately circular unresponsive spot at pixel coordinates
(274,1642), $\approx$45~pixels in diameter, and bad
columns beginning at (1624,2622), (1362,1232), and (546,1468).
This type of SITe CCD is known to generate
spurious charge and exhibit poor CTE performance with too fast parallel
readout, which is why the chip must be clocked rather slowly. As a compromise,
our ST002A is operated with a parallel clock cycle of 800$\mu$sec/phase, which still
generates a spurious charge pattern at a rate of $\approx$3$\times10^{-3}$~e$^-$/row. 
The effect is seen
e.g.\ in bias frames as a vertical ramp with an amplitude of 12~e$^-$/pixel,
adding a noise pedestal of 3~e$^-$ in the uppermost rows (and intermediate
values below). The chip exhibits another disturbing feature which was initially
seen in bias frames as a bright vertical bar of width $\approx$1100~columns
and with an amplitude of 9~e$^-$/pixel. This defect, however, disappeared
when optimizing the clock pattern during wipe cycles such as to direct
charge transfer {\em away} from the serial register rather than conventionally
{\em into} it (reverse clocking). The quantum efficiency of the spectrograph camera (1)
CCD was measured to be 46\%, 65\%, and 28\%, at the wavelengths of 400nm, 650nm,
and 900nm,  respectively (albeit with the caveat 
of not having double-checked our laboratory flux standard against an
independent secondary calibrator).

The mosaic spectrograph camera (2) is combining a science grade ST002A (ron=3.8~e$^-$,
k=1.36~e$^-$/ADU) with an engineering grade chip (ron=9.7~e$^-$,
k=1.60~e$^-$/ADU). The former chip has developed a most unfortunate
luminescence effect as described by Janesick (2001), which has prevented us 
from combining it with the functional CCD of camera (1), see Roth et al.~2004c
for details. Further efforts in optimizing the clock voltages (held in
inversion during exposures) have been successful in so far as the luminescence
effect has vanished. A residual hot column pattern, whose charge
level is extremely sensitive to temperature changes, has rendered the mosaic
camera unsuited for faint object spectroscopy, however still an option for
bright targets where wavelength coverage is an issue.

\subsection{A\&G CCD photometry and astrometry}

Although the A\&G CCD camera was not designed to provide precision photometry 
in the first place, it is nonetheless considered a potentially useful
calibrator for differential spectrophotometry, i.e.\ correcting for
non-photometric conditions due to thin clouds as put forward e.g.\ by 
Barwig et al.\ 1987 (for a first application with PMAS, see Christensen
et al.\ 2003). With observations of a stellar sequence in field
F1038-8 from Stobie et al.\ 1985, we calibrated the A\&G CCD camera
with a series of 4 exposures of 30~s each, obtained on August 28, 2002,
at an airmass of 1.17$\ldots1.20$. We measured 
Star~1 (V=12.99), Star~2 (V=14.28), Star~4 (V=17.94), Star~5 (V=18.25),
and Star~9 (V=20.25), which are simultaneously visible in the FOV.
The exposures were taken through the PMAS V band filter, which is
built according to the recipe by Bessell et al.\ 1990. Excluding Star~9
whose photometry is background-limited and thus less suitable for
calibration, we fitted the count rates CTS [ADU/s] from aperture photometry of the
4 frames and obtained mag~=~-2.5$\;$log~(CTS) + ZP, with a zero point
ZP~=~-24.225~$\pm$~0.02. The 2\% error reflects the slightly non-photometric
conditions of the night.

In order to check the optical system of the A\&G Camera system for
obvious distortions, archival data from an observing run in September
2002 were analyzed, using full-frame acquisition frames from 3
different fields. The observations were acquired in the PMAS
V and R filters, yielding identical results. The astrometry was
performed using the Starlink GAIA tool, comparing star detections
against DSS plates and the USNO catalog. For each frame, 14-17 stars
were found and their astrometric solution determined by the programme.
As a result, the pixel scale was determined to
0.1966$\pm$0.0004~arcsec/pixel. The result is in very good agreement
with the design value of 2.0~arcsec/pixel. The orientation, whose precise 
value depends on the position of the Cassegrain flange rotator after
instrument changes, was determined to 90.831$\pm$0.042 degrees.
In practical terms, N appears to the left, E to the bottom, when
using conventional display tools with CCD pixel~(0,0) at the lower left.

\subsection{Operation at the CAHA 3.5m Telescope}
\label{sec:CAHA}

From the  Spring Semester 2001 to Fall 2004, PMAS was scheduled for 
33 observing runs over a total of 145 nights, and an additional number 
of 27 Service A buffer nights for several high-ranked  programmes.
Throughout these observing runs, PMAS worked reliably
without any major failure, and not a single night was lost due to
technical problems. Some selected science results from these first years
of operation include the following: complete characterization of
PSF as sampled by IFU, yielding a centroiding accuracy at a level
of 10$^{-3}$~arcsec, application to accurate spectrophotometry of faint, 
background-limited Planetary Nebulae in the bulge of M31 (Roth et al.\ 2004b),
PSF-fitting analysis of multiply lensed QSOs (Wisotzki et al. 2003, 2004), 
first 3D spectrophotometry of a supernova type Ia (Christensen et al.\ 2003),
3D spectroscopy of Ly-$\alpha$ emitters associated with the DLA system
Q2233+131 (Christensen et al.\ 2004), abundances and kinematics of a candidate 
sub-damped Ly$_\alpha$ galaxy toward PHL 1226 (Christensen et al.\ 2005),
the ultra-luminous Xray source X-1 in Holmberg~II (Lehmann et al.\ 2004),
crowded field 3D spectroscopy on LBV candidate and circumstellar nebula
in M33 (Becker et al.\ 2004), and 
others \footnote{\url{see http://www.aip.de/groups/opti/pmas/OptI\_pmas.html}}.
PMAS has been offered as a common user instrument under contract with
MPIA Heidelberg from the F2002 semester for a period of 3 years and
possible future extensions. During this period of operation, PMAS
has enjoyed considerable interest from observers to the extent that it 
performed as the 2nd most demanded instrument at the 3.5m Telescope over the
past two years.

\section{Summary}
PMAS is the first dedicated 3D Spectrophotometer with low to medium
spectral resolution, showing a peak instrumental transmission of 24\% and
good performance over the whole optical window 0.35---1~$\mu$m,
especially near the atmospheric cutoff in the blue. The internal
accuracy for flux calibrations over several hours is limited to 
$\approx$1\%. The concept of a 
fiber-coupled lens array type of IFU
has proven to be capable of accurately recording the PSF with a centroiding
accuracy at the milli-arcsec level. PMAS has been used successfully for a 
variety of astrophysical problems, yielding a number of original publications
which have indeed benefitted from high spatial resolution, 3D spectrophotometry,
and the additional information one can retrieve from PSF fitting techniques.
The instrument has been reliably working over 3 years without any major
failure. Newly installed features include PPak, the currently largest FOV IFU
world-wide, a nod-shuffle mode of operation for accurate sky-subtraction,
and the PYTHEAS mode for an increased spectral resolution over a wide
free spectral range. Two underperforming properties viz.\ (1) flexure, introduced
by the unstable focusing stages of the FSPEC optical systems, and 
(2) rather modest QE and overall quality of both the single CCD and the mosaic CCD
cameras, are planned to be fixed in a future overhaul and detector upgrade, respectively.
The successful exploration of PSF-fitting techniques and crowded-field
3D spectroscopy has important implications for a new generation of
advanced 3D spectrographs such as MUSE for the ESO-VTL (Bacon et al.\ 2004).

\vskip 5mm
\centerline{ACKNOWLEDGMENTS}

      Part of this work was supported by the German
      \emph{Deut\-sche For\-schungs\-ge\-mein\-schaft, DFG\/} under
      grant HA1850/10-3 , and by the German Ver\-bund\-for\-schung 
      des BMBF, grants 053PA414/1,  05AL9BA1/9, and 05AE2BAA/4.
      MMR, TB, PB, and AK acknowledge DFG travel grants for several
      observing runs at Calar Alto.
      MMR, TB and JS are grateful to Victor Afanasiev and Serguei 
      Dodonov, Special Astrophysical Observatory in Selentchuk, Russia,
      for hospitality during observing runs at the 6m BTA, and for 
      fruitful discussions and insight into 3D spectroscopy. 
      The authors wish to thank ESO ODT and former team leader Jim Beletic 
      for permission to copy the ESO design CCD detector head and 
      the ACE CCD controller, as well as advice from Roland Reiss, Sebastian
      Deiries, and generous support throughout.
      The support of Calar Alto staff during the PMAS commissioning 
      phase and in operation is gratefully acknowledged. Thanks are  
      due to Sebastian Sanchez for help with grating efficiency measurements,
      and for designing the PMAS archive database. 
      The PMAS Team is especially indebted to Nicolas Cardiel for excellent 
      support during normal and service mode observations.

\clearpage

\end{document}